\documentclass[aps,twocolumn,floatfix,superscriptaddress,preprintnumbers,showpacs,showkeys]{revtex4}
\usepackage{amssymb}
\usepackage{amsmath}
\usepackage{amsfonts}
\usepackage{epsfig}
\usepackage{graphicx}
\usepackage{tabularx}
\newcommand{\seq}{\begin{subequations}}
\newcommand{\sen}{\end{subequations}}
\newcommand{\eq}{\begin{eqnarray}}
\newcommand{\en}{\end{eqnarray}}

\def\shiftdown#1{#1\llap{\lower.04ex\hbox{#1}}}

\newcommand{\ra}{\rangle}
\newcommand{\la}{\langle}

\newcommand{\bfq}{{\bf q}_{\perp}}

\newcommand{\bfk}{{\bf k}_{\perp}}

\begin{document}

\title{Nucleon structure in a light-front quark model \\ 
consistent with quark counting rules and data}

\author{Thomas Gutsche}
\affiliation{Institut f\"ur Theoretische Physik,
Universit\"at T\"ubingen, 
Kepler Center for Astro and Particle Physics,  
Auf der Morgenstelle 14, D-72076 T\"ubingen, Germany}
\author{Valery E. Lyubovitskij} 
\affiliation{Institut f\"ur Theoretische Physik,
Universit\"at T\"ubingen, 
Kepler Center for Astro and Particle Physics,  
Auf der Morgenstelle 14, D-72076 T\"ubingen, Germany}
\affiliation{Department of Physics, Tomsk State University,  
634050 Tomsk, Russia} 
\affiliation{Mathematical Physics Department, 
Tomsk Polytechnic University, 
Lenin Avenue 30, 634050 Tomsk, Russia} 
\author{Ivan Schmidt}
\affiliation{Departamento de F\'\i sica y Centro Cient\'\i fico 
Tecnol\'ogico de Valpara\'\i so (CCTVal), Universidad T\'ecnica
Federico Santa Mar\'\i a, Casilla 110-V, Valpara\'\i so, Chile}
\author{Alfredo Vega}
\affiliation{Instituto de F\'isica y Astronom\'ia, 
Universidad de Valpara\'iso,  
Avenida Gran Breta\~na 1111, Valpara\'iso, Chile}
\affiliation{Centro de Astrof\'isica de Valpara\'iso, 
Avenida Gran Breta\~na 1111, Valpara\'iso, Chile} 

\date{\today}

\begin{abstract}

Using global fits of valence $u$ and $d$ quark parton distributions 
and data on quark and nucleon form factors in the Euclidean region, 
we derive a light-front quark model for the nucleon structure  
consistent with quark counting rules. 

\end{abstract}



\maketitle

\section{Introduction}

The main objective of this paper is to continue our study of  
a phenomenological light-front wave function (LFWF) for the 
nucleon started as in Ref.~\cite{Gutsche:2013zia}. 
We derived a LFWF for hadrons for both pions and nucleons  
which at an initial scale is constrained by the soft-wall 
anti-de Sitter (AdS)/QCD 
model, and which at higher scales gives the correct scaling behavior 
of parton distributions and form factors.
The explicit form of the wave function at large scales is extracted 
from the hard evolution of parton distribution functions (PDFs) and 
generalized parton distributions (GPDs).  
The proposed wave function produces form factors consistent with quark 
counting rules~\cite{Brodsky:1973kr} 
and also gives predictions for the corresponding parton 
distributions. In our considerations we obtained harder PDFs in comparison 
with the results of global fits [see e.g. results of Martin, Stirling, 
Thorne and Watt (MSTW)~\cite{Martin:2009iq}]. The reason for a softening 
of the PDFs was discussed in the pion case in Ref.~\cite{Aicher:2010cb}. 
There it was clearly demonstrated that the inclusion of
next-to-leading logarithmic threshold resummation effects, 
due to collinear and soft gluon contributions, leads to a softer 
pion PDF~\cite{Aicher:2010cb}. 
This result also shows that we should take into account these resummation 
effects and derive an improved nucleon LFWF. In Ref.~\cite{Gutsche:2014zua}  
we demonstrate how to derive in the case of the pion a LFWF producing 
a softer PDF as in Ref.~\cite{Aicher:2010cb} and a pionic electromagnetic 
form factor consistent with data and quark counting rules. 
Here we extend this idea to the case of the nucleon. 
We propose a LFWF for the nucleon modeled as a quark-scalar diquark 
bound state, with a specific dependence on the transverse momentum $\bfk$ 
and the light-cone variable $x$. This LFWF produces PDFs for the valence 
$u$ and $d$ quarky found in the global fits of Ref.~\cite{Martin:2009iq}. 
It also describes the electromagnetic form factors of the nucleon 
including their flavor decomposition into $u$ and $d$ quark form factors 
up to values of the momentum transfer squared $Q^2 = 30$ GeV$^2$ 
in the Euclidean region (for a recent overview of experimental and 
theoretical progress in the study of nucleon electromagnetic structure 
see e.g. Refs.~\cite{Cates:2011pz}-\cite{Obukhovsky:2013fpa}).  
It is important to stress that the calculated nucleon electromagnetic 
form factors are consistent with quark counting rules 
for large values~of~$Q^2$.  

The main advantage of our approach is that the derived LFWF does not depend
on phenomenological parameters like masses of quark/diquark,
which are not directly related to QCD. Restricting to zero
current quark masses we obtain a reasonable description of
data on nucleon form factors.
Note, we derive the LFWF at the initial scale $\mu_0 = 1$ GeV
producing corresponding PDFs at the same scale. Both quantities
are scale dependent. We showed explicitly
in our paper~\cite{Gutsche:2013zia} that at 
any scale evolved PDF could constraint the corresponding LFWF. 
The most convenient way is to set up LFWF at the initial scale,
calculate the corresponding PDF and then perform an evolution of 
the PDF to higher scales. On the other hand, as we stressed before, 
we make a further improvement on the light-front quark model for 
nucleon proposed in 
Ref.~\cite{Gutsche:2013zia} producing softer quark PDFs 
at the initial scale $\mu_0 = 1$ GeV in agreement with global 
analysis of these quantities from the data (see details in Sec.II).  
The description of nucleon structure starting from light-front wave 
functions is also our improvement in comparison with other approaches 
calculating  nucleon form factors using parametrization for 
generalized parton distributions~\cite{Diehl:2004cx}. 

\section{Light-front quark-diquark model for the nucleon} 

In this section we propose a phenomenological LFWF 
$\psi(x,\bfk)$ for the nucleon, set up as a bound state 
of an active quark and a spectator scalar diquark. This LFWF is able 
to produce the $u$ and $d$ quark PDFs derived in the 
global fits of Ref.~\cite{Martin:2009iq} and generates electromagnetic
form factors of nucleons including their flavor decomposition 
which are consistent with data.  

First we collect the well-known decompositions~\cite{Radyushkin:1998rt} 
of the nucleon Dirac and Pauli form factors $F_{1,2}^N$ ($N=p,n$) in terms of
the valence quark distributions in nucleons with $F_{1,2}^q$ ($q=u,d$), 
which then are related to the GPDs ($\mathcal{H}^{q}$ and 
$\mathcal{E}^{q}$)~\cite{Mueller:1998fv} of valence quarks 
\eq\label{FF}
F_i^{p(n)}(Q^2) &=& \frac{2}{3} F_i^{u(d)}(Q^2) 
                 -  \frac{1}{3} F_i^{d(u)}(Q^2)\,, 
\nonumber\\
F_1^q(Q^2) &=& \int_{0}^{1} dx \, \mathcal{H}^{q}(x,Q^2)  \,,\\
F_2^q(Q^2) &=& \int_{0}^{1} dx \, \mathcal{E}^{q}(x,Q^2)  \, .
\nonumber 
\en 
At $Q^2=0$ the GPDs are related to the quark densities --- 
valence $q_v(x)$ and magnetic $\mathcal{E}_{q}(x)$ as 
\eq\label{norm_GPDs}
\mathcal{H}^{q}(x,0)=q_{v}(x)\,, 
\quad 
\mathcal{E}^{q}(x,0)=\mathcal{E}^{q}(x) \,, 
\en
which are normalized as 
\eq\label{normalization} 
n_q &=& F_1^q(0) = \int\limits_0^1 dx \, q_v(x)\,, \nonumber\\
\kappa_q &=& F_2^q(0) = \int\limits_0^{1} dx \, \mathcal{E}^q(x)\, . 
\en
The number of $u$ or $d$ valence quarks in the proton is denoted
by $n_q$, and $\kappa_q$ is the quark anomalous magnetic moment. 

Next we recall the definitions of the nucleon Sachs form factors 
$G_{E/M}^N(Q^2)$ and the electromagnetic radii $\la r^2_{E/M} \ra^N$ 
in terms of the Dirac and Pauli form factors 
\eq 
G_E^N(Q^2) &=& F_1^N(Q^2) - \frac{Q^2}{4m_N^2} F_2^N(Q^2)\,, 
\nonumber\\
G_M^N(Q^2) &=& F_1^N(Q^2) + F_2^N(Q^2)\,, \nonumber\\[2mm]
\la r^2_E \ra^N &=& - 6 \, \frac{dG_E^N(Q^2)}{dQ^2}\bigg|_{Q^2 = 0} \,, 
\nonumber\\
\la r^2_M \ra^N &=& - \frac{6}{G_M^N(0)} \, 
\frac{dG_M^N(Q^2)}{dQ^2}\bigg|_{Q^2 = 0}  \,, 
\en 
where $G_M^N(0) \equiv \mu_N$ is the nucleon magnetic moment. 

The light-front representation~\cite{Brodsky_Drell,LFQCD}  
for the Dirac and Pauli quark form factors is  
\eq 
F_1^q(Q^2) &=& 
\int\limits_0^1 dx 
\int\frac{d^2\bfk}{16\pi^3} \, 
\biggl[ 
\psi_{+q}^{+\, \ast}(x,\bfk')\psi_{+q}^+(x,\bfk) 
\label{FF_LFQCD1} \nonumber\\
&+&\psi_{-q}^{+\, \ast}(x,\bfk')\psi_{-q}^+(x,\bfk) 
\biggr] \,, \\
F_2^q(Q^2) &=& - \frac{2M_N}{q^1-iq^2}
\int\limits_0^1 dx 
\int\frac{d^2\bfk}{16\pi^3} \nonumber\\
&\times&
\biggl[ 
\psi_{+q}^{+\, \ast}(x,\bfk')\psi_{+q}^-(x,\bfk) 
\nonumber\\
&+&\psi_{-q}^{+\, \ast}(x,\bfk')\psi_{-q}^-(x,\bfk) 
\biggr] \,. \label{FF_LFQCD2} 
\en  
Here $M_N$ is the nucleon mass, $\psi_{\lambda_q q}^{\lambda_N}(x,\bfk)$ 
are the LFWFs at the initial scale $\mu_0$ with specific helicities for  
the nucleon $\lambda_N  = \pm$ and for the struck quark $\lambda_q = \pm $, 
where plus and minus correspond to $+\frac{1}{2}$ and $-\frac{1}{2}$, 
respectively. We work in the frame with $q=(0,0,\bfq)$, and where  
the Euclidean momentum squared is $Q^2 = \bfq^2$. As the initial scale 
we choose the value $\mu_0 = 1$ GeV which is used in the MSTW 
global fit~\cite{Martin:2009iq}. 

In the quark-scalar diquark model, the generic ansatz for the massless 
LFWFs at the initial scale $\mu_0 = 1$ GeV reads 
\eq 
\psi_{+q}^+(x,\bfk) &=& \varphi_q^{(1)}(x,\bfk) 
\,, \nonumber\\
\psi_{-q}^+(x,\bfk) &=& -\frac{k^1 + ik^2}{xM_N} 
\, \varphi_q^{(2)}(x,\bfk)  \,, \nonumber\\[2mm]
\psi_{+q}^-(x,\bfk) &=& \frac{k^1 - ik^2}{xM_N} 
\, \varphi_q^{(2)}(x,\bfk)  
\,, \nonumber\\
\psi_{-q}^-(x,\bfk) &=& \varphi_q^{(1)}(x,\bfk) 
\,, 
\en 
where 
$\varphi_q^{(1)}$ and $\varphi_q^{(2)}$ are 
the twist-3 LFWFs. They are generalizations of the twist-3 LFWFs 
found from matching the electromagnetic form factors of the nucleon in 
soft-wall AdS/QCD~\cite{Brodsky:2007hb}-\cite{Gutsche:2011vb} and 
light-front QCD (see the detailed discussion in Ref.~\cite{Gutsche:2013zia}).  
In particular, as a result of the matching the following LFWFs 
have been deduced:  
\eq 
\varphi_q^{{\rm AdS/QCD} (i)}(x,\bfk) &=& 
N_{q}^{(i)} \, \frac{4\pi}{\kappa} \, 
\sqrt{\frac{\log(1/x)}{1-x}} \nonumber\\ 
&\times&\exp\biggl[- \frac{\bfk^2}{2\kappa^2}
\, \frac{\log(1/x)}{(1-x)^2}
\biggr] \,, 
\en 
where the $N_{q}^{(i)}$ are normalization constants fixed by 
the conditions of~(\ref{normalization}). 
Note that the derived LFWF is not symmetric under the exchange 
$x \to 1-x$. This is the case because it was extracted from a matching 
of matrix elements of the bare electromagnetic current between 
the dressed LFWF in light-front QCD and of the
dressed electromagnetic current between hadronic wave 
functions in AdS/QCD. 

The generalization 
$\varphi_q^{{\rm AdS/QCD} (i)}(x,\bfk) \to \varphi_q^{(i)}(x,\bfk)$ 
is encoded in the longitudinal factors $f_q^{(i)}(x)$ and $\bar f_q(x)$  
which take into account collinear and soft gluon effects as 
\eq 
\varphi_q^{(i)}(x,\bfk) &=&  
N_{q}^{(i)} \, \frac{4\pi}{\kappa} \, 
\sqrt{\frac{\log(1/x)}{1-x}} \, \sqrt{f_q^{(i)}(x) \bar f_q(x)}
\nonumber\\
&\times&\exp\biggl[- \frac{\bfk^2}{2\kappa^2}
\, \frac{\log(1/x)}{(1-x)^2} \, \bar f_q(x) \biggr] \,. 
\en 
These factors lead to softer PDFs, which coincide 
with the results of the global fit performed e.g. 
in Ref.~\cite{Martin:2009iq}. At the same time, the power scaling 
of electromagnetic form factors for large values of Euclidean momentum 
squared with $Q^2 \to \infty$ remains the same up to 
power-scaling breaking corrections $\Delta_q^{(i)}$ 
[see Eqs.~(\ref{Delta_q1}) and~(\ref{Delta_q2})], which produce fine-tuned  
fits of the nucleon electromagnetic form factors, i.e. consistent with 
quark counting rules. The choice of the functions $f_q^{(i)}(x)$ 
is constrained by the valence $u$ and $d$ quark PDF, while $\bar f_q(x)$ 
is fixed from the fit to quark and nucleon form factors.  
The functions $f_q^{(i)}(x)$ and $\bar f_q(x)$ are specified as 
\eq 
f_q^{(1)}(x) &=& x^{\eta^{(1)}_q-1} \, (1-x)^{\eta^{(2)}_q-1} \, 
(1 + \epsilon_q \sqrt{x} + \gamma_q x) \,, 
\nonumber\\
f_q^{(2)}(x) &=& x^{2+\rho_q} (1-x)^{\sigma_q} \, 
(1 + \lambda_q \sqrt{x} + \delta_q x)^2 \, 
f_q^{(1)}(x) \,, 
\nonumber\\
\bar f_q(x) &=& 
x^{\bar\eta^{(1)}_q} \, (1-x)^{\bar\eta^{(2)}_q} \, 
(1 + \bar\epsilon_q \sqrt{x} + \bar\gamma_q x) \,, 
\en 
where the parameters $\eta^{(1)}_q$, $\eta^{(2)}_q$, 
$\epsilon_q$ and $\gamma_q$ are fixed from the global 
MSTW analysis of~Ref.~\cite{Martin:2009iq} (for simplicity 
we restrict ourselves to leading-order results). The parameters $\rho_q$, 
$\sigma_q$, $\lambda_q$, $\delta_q$, 
$\bar\eta^{(1)}_q$, $\bar\eta^{(2)}_q$, 
$\bar\epsilon_q$ and $\bar\gamma_q$ are fixed from 
a fit to the anomalous magnetic moments of quarks (nucleons) 
and to the $Q^2$ dependence of the electromagnetic quark (nucleon) 
form factors. The final set of parameters specifying the functions 
$f_q^{(i)}(x)$ and $\bar f_q(x)$ is listed in Table~I. 

\begin{table}[htb]
\begin{center}
\caption{Parameters specifying $f_q^{(i)}(x)$ and $\bar f_q(x)$}

\vspace*{.1cm}

\def\arraystretch{1.2}
\hspace*{-.4cm}
\begin{tabular}{|c|c||c|c|}
\hline
Parameter   &  Value  &  Parameter   &  Value \\
\hline
$\eta^{(1)}_u$    & 0.45232  & $\eta^{(1)}_d$    & 0.71978  \\
$\eta^{(2)}_u$    & 3.0409   & $\eta^{(2)}_d$    & 5.3444   \\
$\epsilon_u$  & $-2.3737$  & $\epsilon_d$  & $-4.3654$      \\
$\gamma_u$    & 8.9924   & $\gamma_d$    & 7.4730           \\
$\bar\eta^{(1)}_u$& 0.195    & $\bar\eta^{(1)}_d$& 0.280    \\
$\bar\eta^{(2)}_u$& $\frac{\eta^{(2)}_u - 1}{2} - 0.54$ &
$\bar\eta^{(2)}_d$& $\frac{\eta^{(2)}_d - 1}{2} - 0.60$   \\
$\bar\epsilon_u$  & $-0.71$  & $\bar\epsilon_d$  & $-0.10$    \\
$\bar\gamma_u$    &  0     & $\bar\gamma_d$      &  0     \\ 
$\rho_u$          &  0.091 & $\rho_d$          & $-0.17$    \\ 
$\sigma_u$        &  $(\eta^{(2)}_u - 1) - 0.2409$    & 
$\sigma_d$        &  $(\eta^{(2)}_d - 1) - 2.3444$        \\
$\lambda_u$       & $-2.40$   & $\lambda_d$      & $-0.22$    \\ 
$\delta_u$        &  3.18  & $\delta_d$        &  3.90    \\ 
\hline
\end{tabular}
\end{center}
\end{table} 
The scale parameter $\kappa = 350$ MeV remains the same 
as fixed in the analysis of Ref.~\cite{Gutsche:2013zia} 
and used in the analysis for the pion of Ref.~\cite{Gutsche:2014zua}. 
The parameter $\kappa$ is related to the scale parameter of 
the background dilaton field providing confinement and 
is universal for all hadronic wave functions. 

The expressions for the quark PDFs read
\eq 
q_v(x) &=& (N_q^{(1)})^2 (1-x) f_q^{(1)}(x) \nonumber\\
       &+& (N_q^{(2)})^2 \frac{\kappa^2}{M_N^2} 
\frac{(1-x)^3}{x^2 \log(1/x)} \, \frac{f_q^{(2)}(x)}{\bar f_q(x)}\,, 
\label{qvx}\\
\mathcal{E}^q(x) &=& 2 N_q^{(1)} N_q^{(2)} 
\frac{(1-x)^2}{x} \,
\sqrt{f_q^{(1)}(x) f_q^{(2)}(x)}
\label{Evx} \,. 
\en 
The ratio $c_q = N_q^{(2)}/N_q^{(1)}$ is a free parameter and we choose 
for simplicity $c_u = 1$ and $c_d = -1$ or 
$N_u^{(1)} = N_u^{(2)} = N_u$ and $N_d^{(1)} = - N_d^{(2)} = N_d$. 
In our calculations normalization constants $N_u$ and $N_d$ are fixed as 
$N_u = 1.18093$ and $N_d = 2.00432$. 
Notice that the contribution of the struck quark with negative helicity 
$\lambda_q = -$ [see second term in Eq.~(\ref{qvx})] to 
the quark PDFs $q_v(x)$ is relatively suppressed by a factor 
$\kappa^2/M_N^2 \sim 1/10$. To match the $u_v(x)$ and $d_v(x)$ 
PDFs fixed in the global fit of Ref.~\cite{Martin:2009iq}, we 
slightly change the parameters in the longitudinal factor $f_q(x)$. 
We found that in the case of the PDF $u_v(x)$ the contribution 
of the struck quark with negative helicity is negligible. 
In the case of the PDF $d_v(x)$ we slightly change 
the parameter $\eta^{(2)}_d = 5.1244$ fixed in Ref.~\cite{Martin:2009iq} 
to $\eta^{(2)}_d = 5.3444$ to match the results of 
the global fit in~\cite{Martin:2009iq}. 

\begin{figure}
\begin{center}

\epsfig{figure=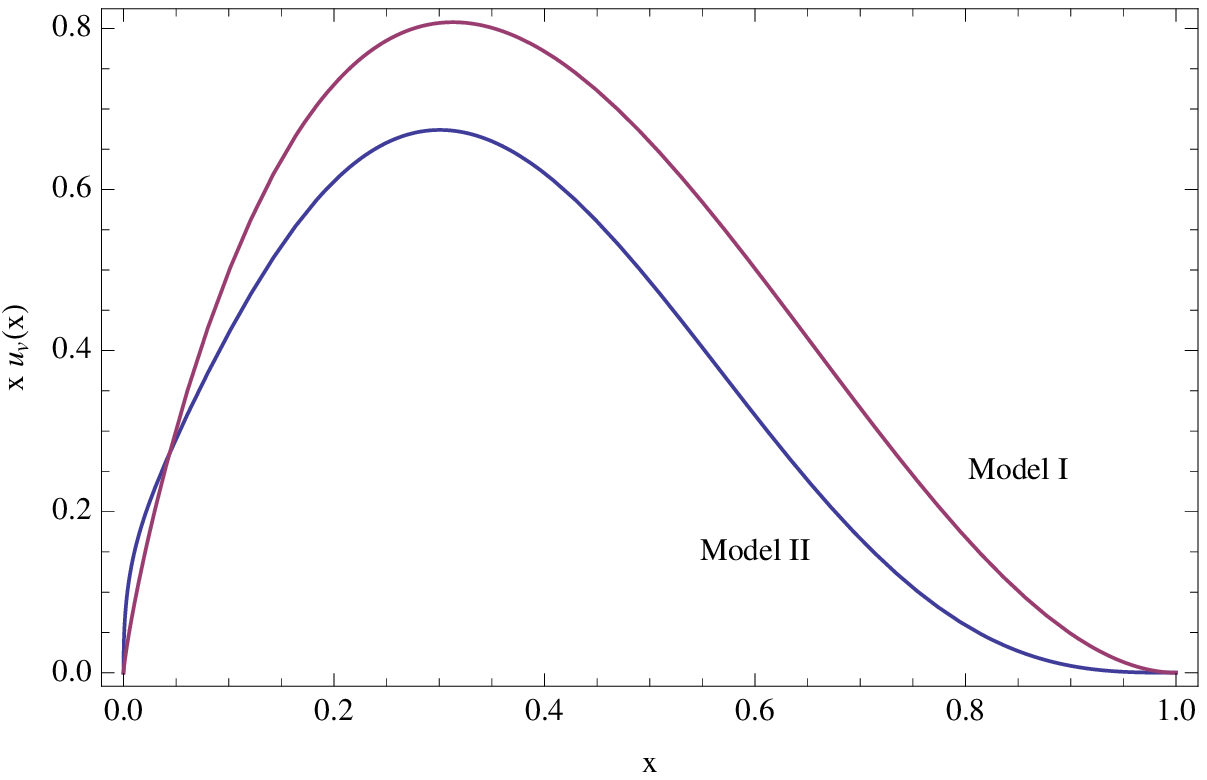,scale=.575}
\end{center}
\vspace*{-.6cm}
\noindent 
\caption{$u_v(x)$ at scale $\mu=1$ GeV 
in Models I and II. 
\label{fig:pdf1}}  

\vspace*{-.35cm}

\begin{center}
\epsfig{figure=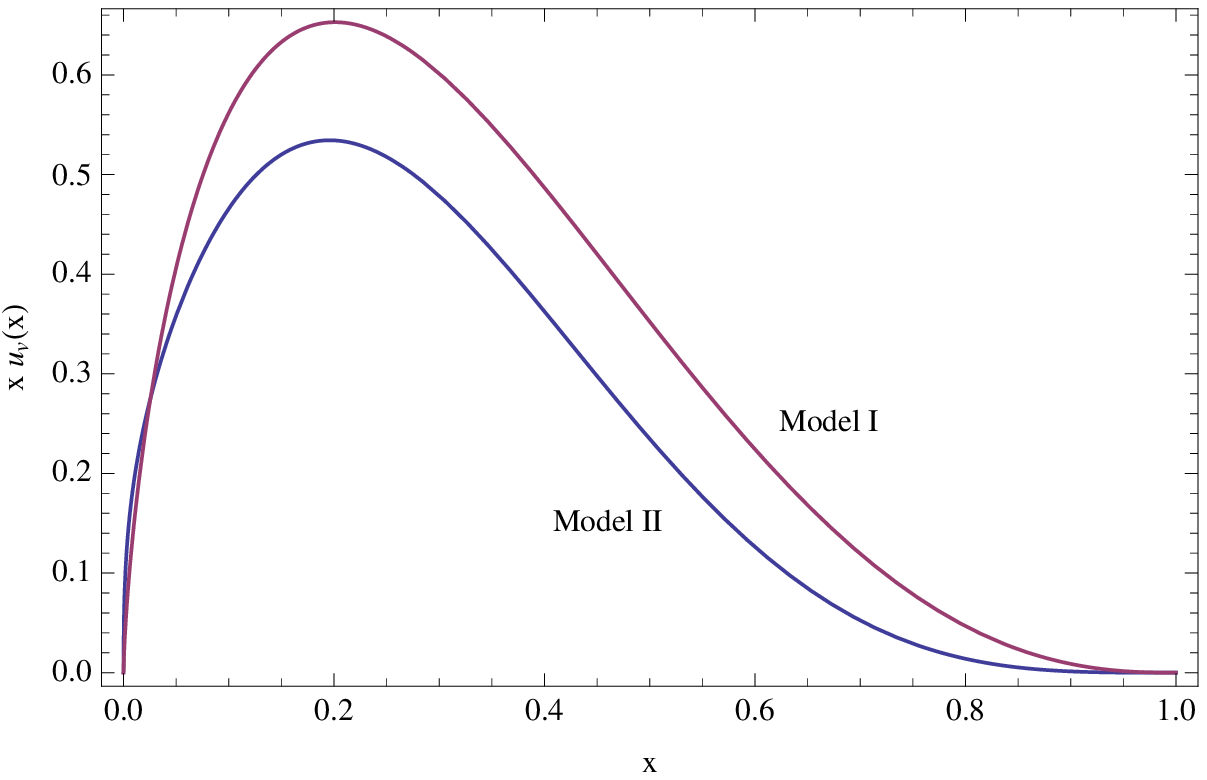,scale=.575}
\end{center}
\vspace*{-.6cm}
\noindent 
\caption{$u_v(x)$ at scale $\mu=10$ GeV 
in Models I and II. 
\label{fig:pdf2}}

\vspace*{-.35cm}

\begin{center}
\epsfig{figure=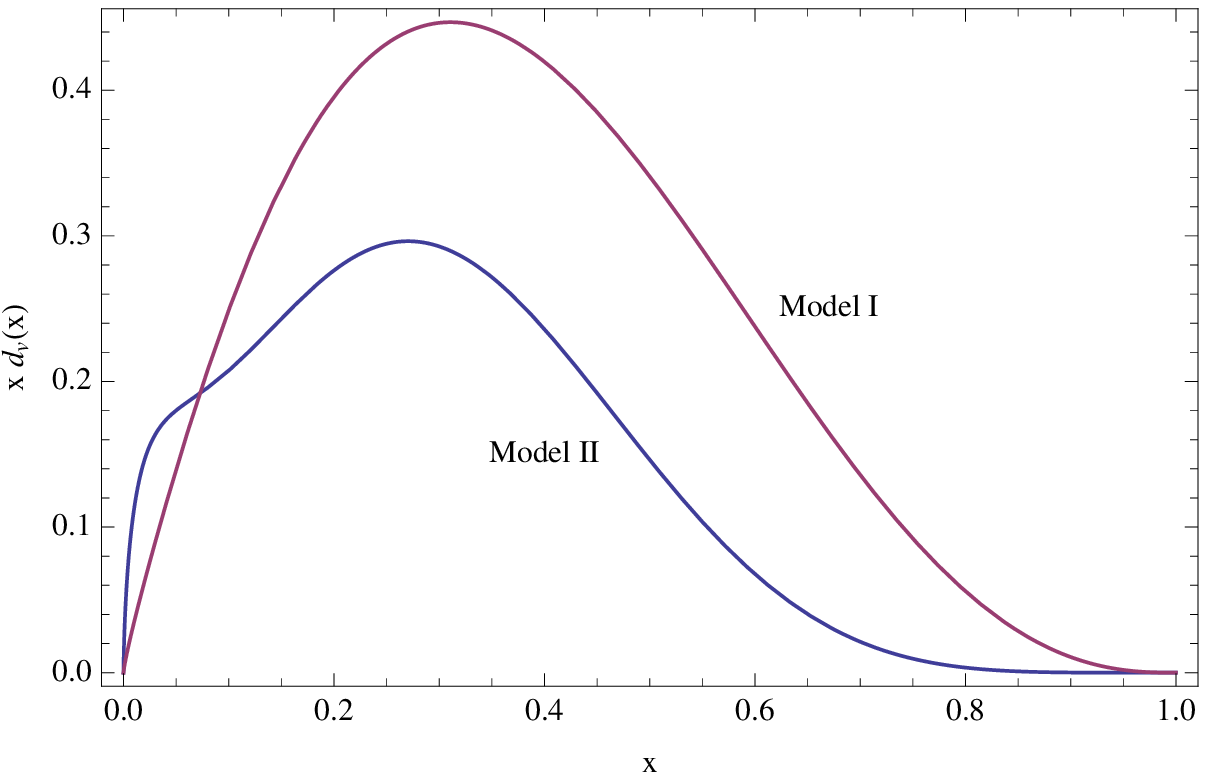,scale=.575}
\end{center}
\vspace*{-.6cm}
\noindent 
\caption{$d_v(x)$ at scale $\mu=1$ GeV 
in Models I and II. 
\label{fig:pdf3}}  

\vspace*{-.35cm}

\begin{center}
\epsfig{figure=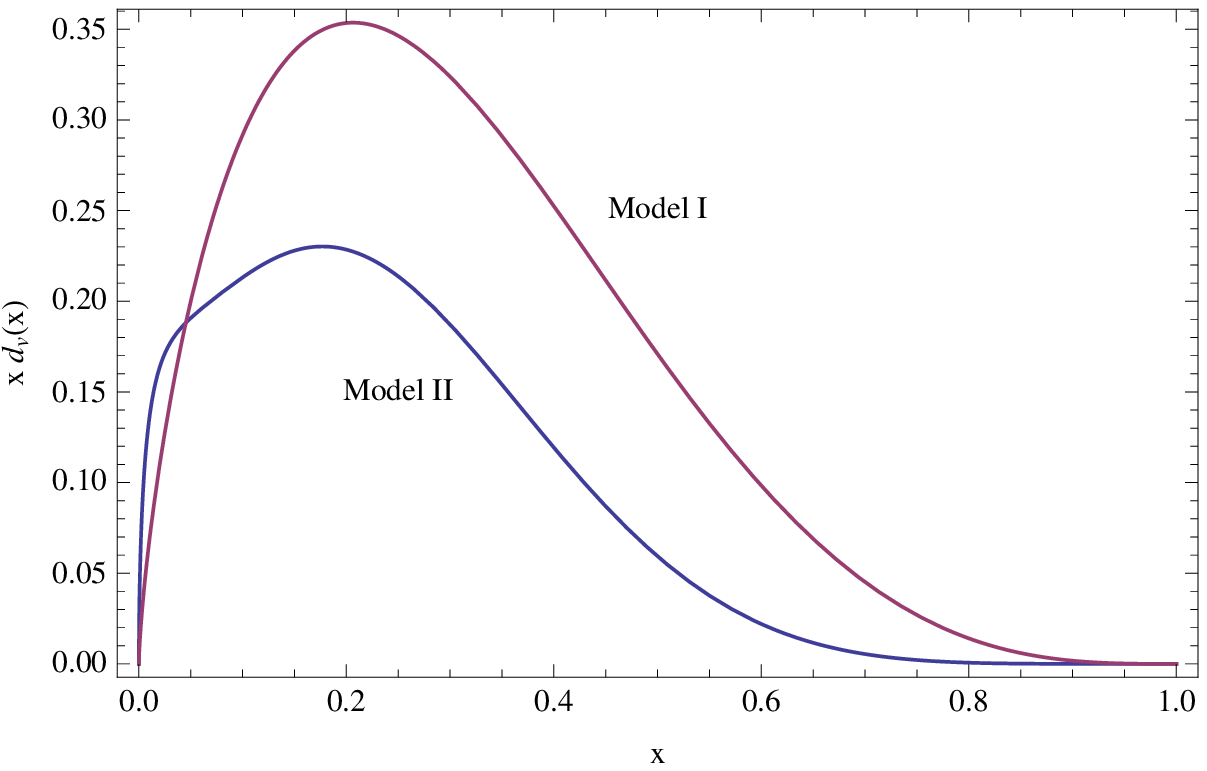,scale=.575}
\end{center}
\vspace*{-.6cm}
\noindent 
\caption{$d_v(x)$ at scale $\mu=10$ GeV 
in Models I and II. 
\label{fig:pdf4}}  
\end{figure}

\begin{figure}
\begin{center}

\epsfig{figure=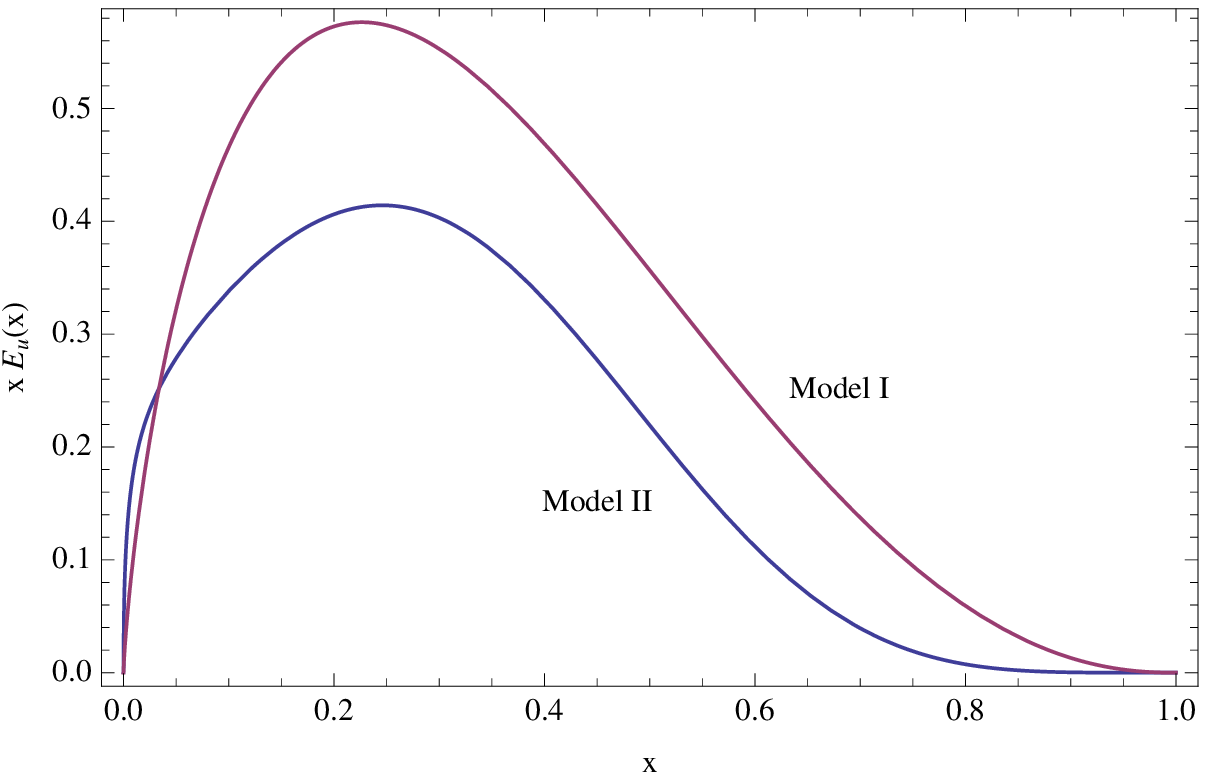,scale=.575}
\end{center}
\vspace*{-.6cm}
\noindent 
\caption{$\mathcal{E}_u(x)$ at scale $\mu=1$ GeV 
in Models I and II. 
\label{fig:pdf5}}  

\vspace*{-.35cm}

\begin{center}
\epsfig{figure=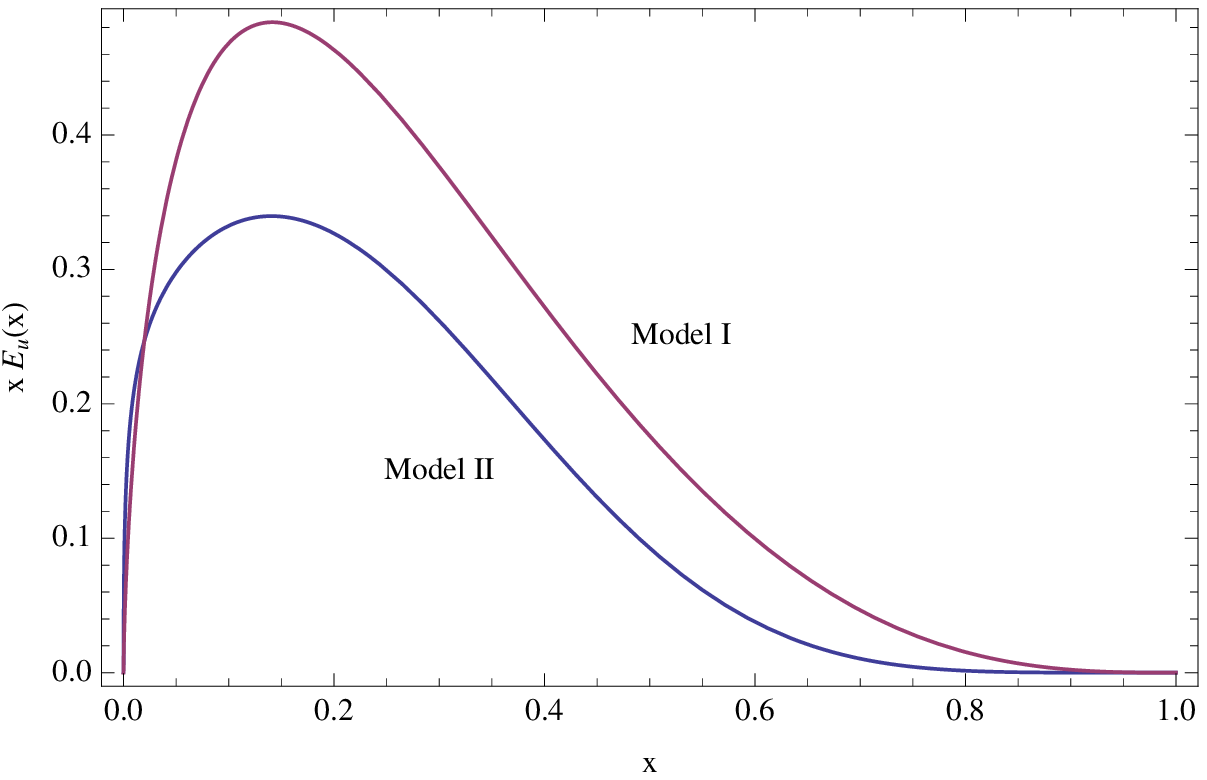,scale=.575}
\end{center}
\vspace*{-.6cm}
\noindent 
\caption{$\mathcal{E}_u(x)$ at scale $\mu=10$ GeV 
in Models I and II. 
\label{fig:pdf6}}

\vspace*{-.35cm}

\begin{center}
\epsfig{figure=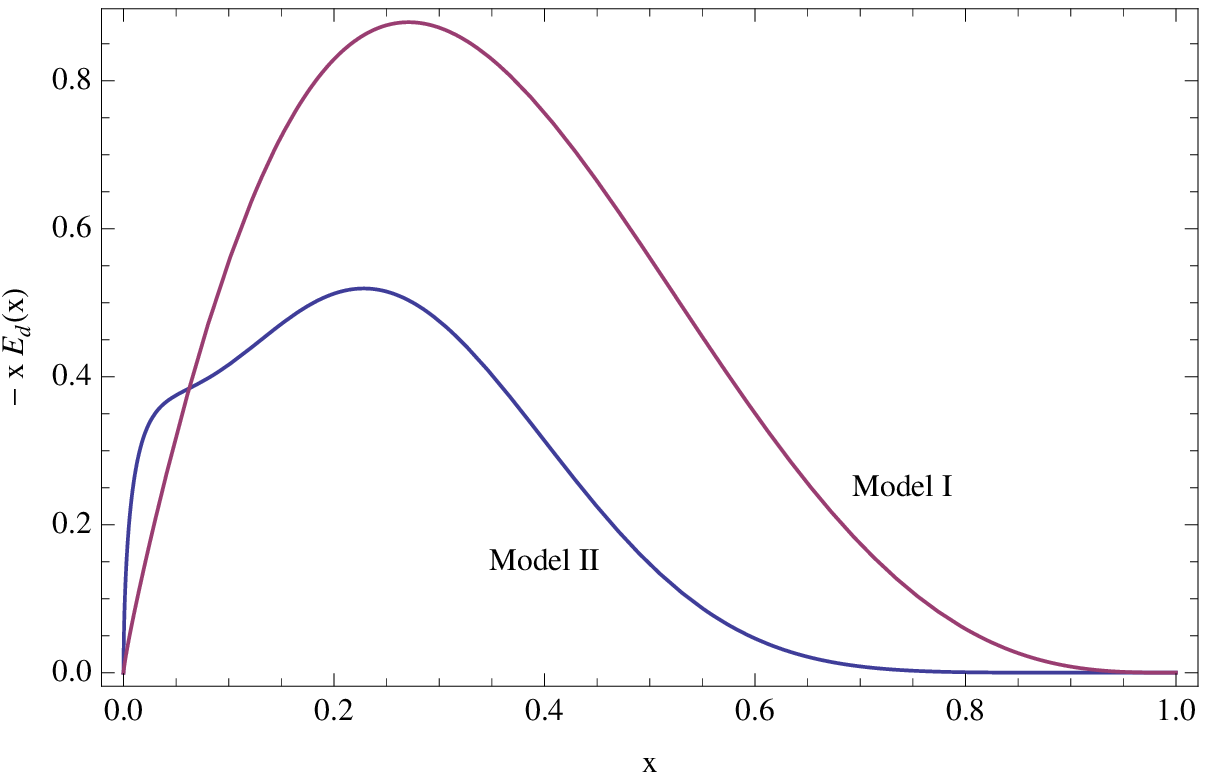,scale=.575}
\end{center}
\vspace*{-.6cm}
\noindent 
\caption{$\mathcal{E}_d(x)$ at scale $\mu=1$ GeV 
in Models I and II. 
\label{fig:pdf7}}  

\vspace*{-.35cm}

\begin{center}
\epsfig{figure=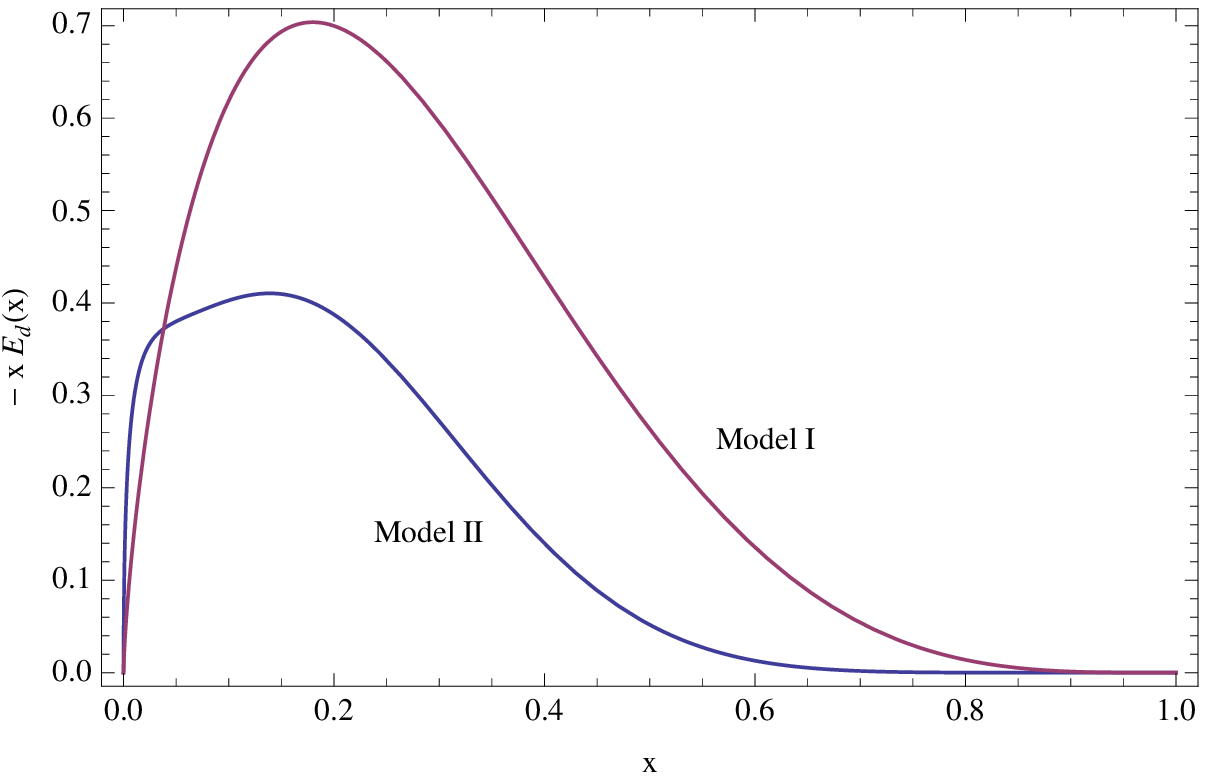,scale=.575}
\end{center}
\vspace*{-.6cm}
\noindent 
\caption{$\mathcal{E}_d(x)$ at scale $\mu=10$ GeV 
in Models I and II. 
\label{fig:pdf8}}  
\end{figure}

Expressions for the quark helicity-independent GPDs $\mathcal{H}^{q}$ 
and $\mathcal{E}^{q}$ in the nucleon read 
\eq 
\mathcal{H}^{q}(x,Q^2) &=& q_v(x,Q^2) 
\exp\biggl[-\frac{Q^2}{4\kappa^2} \, \log(1/x) \bar f_q(x)  \biggr] \,, 
\nonumber\\
q_v(x,Q^2) &=& q_v(x) 
- (N_q^{(2)})^2 \, \frac{Q^2}{4M_N^2} 
\, \frac{(1-x)^3}{x^2} \,  f_q^{(2)}(x) \,, 
\nonumber\\
\mathcal{E}^{q}(x,Q^2) &=& {\cal E}^q(x) 
\exp\biggl[-\frac{Q^2}{4\kappa^2} \, \log(1/x) \bar f_q(x)  \biggr] \,. 
\en 
In the following we consider the scaling of the PDFs at large $x$ and 
form factors at large $Q^2$. The $q_v(x)$ scale at large $x$ as 
\eq 
& &q_v(x) \sim (1-x)^{\eta_q^{(2)}}\,, \nonumber\\
& &u_v(x) \sim (1-x)^{3}\,, \ \ d_v(x) \sim (1-x)^{5}\,. 
\en 
For the $\mathcal{E}^q(x)$ we have the scaling behavior at large $x$ as 
\eq 
& &\mathcal{E}^q(x) \sim q_v(x) \, (1-x)^{1+\sigma_q/2} 
\sim (1-x)^{\eta_q^{(2)}+1+\sigma_q/2}  \,, \nonumber\\
& &\mathcal{E}^u(x)  \sim (1-x)^{5}\,, \ \ 
\mathcal{E}^d(x)  \sim (1-x)^{7}\,. 
\en 
Note that in the case of the $u$ quark PDFs our results are consistent 
with perturbative QCD~\cite{Brodsky:1994kg,Yuan:2003fs}. 
The $d$ quark PDFs have an additional power of 2 in the scaling 
behavior as required by the global analysis. As we mentioned before 
the scaling of the quark PDFs calculated in this paper is softer 
at the initial scale $\mu_0 = 1$ GeV in comparison with results 
of Ref.~\cite{Gutsche:2013zia}, where we considered another version 
of the present light-front quark model. In particular, 
the PDFs calculated in Ref.~\cite{Gutsche:2013zia} scale 
$u_v(x) \sim (1-x)^{2}$\,, $d_v(x) \sim (1-x)^{2.5}$\,, 
$\mathcal{E}^u(x)  \sim (1-x)^{2.3}$\,, and 
$\mathcal{E}^d(x)  \sim (1-x)^{2.8}$\,.  
Note that the quality of the description of nucleon electromagnetic 
form factors is similar in both versions of our approach.   
Therefore, the main advantage of the present version of our approach 
in comparison with the version discussed in Ref.~\cite{Gutsche:2013zia} 
consists in the improvement of the behavior of quark PDFs. 
In Figs.~\ref{fig:pdf1}-\ref{fig:pdf8} 
we perform a comparison of PDFs calculated 
in two versions of our approach (Model I discussed in 
Ref.~\cite{Gutsche:2013zia} and Model II proposed in 
the present paper). For convenience, we present results for the quark PDFs 
at scales 1 and 10 GeV.

In the case of the nucleon form factors we get for large $Q^2$ the behaviors   
\eq
F_1^q(Q^2) &\sim& \int\limits_0^1 dx (1-x)^{\eta_q^{(2)}} \, 
\exp\biggl[-\frac{Q^2}{4\kappa^2} \, (1-x)^{1+\bar\eta_q^{(2)}}
\biggr] \nonumber\\
&\sim& \biggl(\frac{1}{Q^2}\biggr)^{\frac{1+\eta_q^{(2)}}
{1+\bar\eta_q^{(2)}}} 
\sim \biggl(\frac{1}{Q^4}\biggr)^{1+\Delta_q^{(1)}} 
\en
and 
\eq
\hspace*{-.3cm}
F_2^q(Q^2) &\sim& 
\int\limits_0^1 dx (1-x)^{1+\eta_q^{(2)}+\frac{\sigma_q}{2}} 
\exp\biggl[-\frac{Q^2}{4\kappa^2} (1-x)^{1+\bar\eta_q^{(2)}}
\biggr] \nonumber\\
&\sim& \biggl(\frac{1}{Q^2}\biggr)^{\frac{2+\eta_q^{(2)}+\frac{\sigma_q}{2}} 
{1+\bar\eta_q^{(2)}}} 
\sim \biggl(\frac{1}{Q^6}\biggr)^{1+\Delta_q^{(2)}} \,. 
\en
Here, the $\Delta_q^{(1)}$ and $\Delta_q^{(2)}$ are the small corrections 
encoding a deviation of the Dirac and Pauli quark form factors from the 
power-scaling laws $1/Q^4$ and $1/Q^6$, respectively. These corrections 
are given in the form  
\eq\label{Delta_q1} 
\Delta_q^{(1)} &=& 
\frac{1+\eta_q^{(2)}}{2 (1+\bar\eta_q^{(2)})} - 1\,, 
\nonumber\\
\Delta_q^{(2)} &=& \frac{2}{3} \Delta_q^{(1)} +  
\frac{1}{3} \biggl( 
\frac{1+\sigma_q/2}{1+\bar\eta_q^{(2)}} - 1\biggr) 
\en  
and vanish for 
\eq 
\bar\eta_q^{(2)} = \frac{\sigma_q}{2} = 
\frac{\eta^{(2)}_q - 1}{2} \,. 
\en 
The last limit is consistent with the Drell-Yan-West 
duality~\cite{Drell:1969km} relating the large-$Q^2$ behavior of nucleon 
electromagnetic form factors and the large-$x$ 
behavior of the structure functions. 
However, a fine-tuned fit to the electromagnetic form 
factors requires a deviation of $\Delta_q^{(i)}$ from zero with 
the numerical values of  
\eq\label{Delta_q2}  
& &\Delta_u^{(1)} = 0.365\,, \quad 
   \Delta_d^{(1)} = 0.233\,, \nonumber\\
& &\Delta_u^{(2)} = 0.338\,, \quad 
   \Delta_d^{(2)} = 0.081\,. 
\en 

\section{Results}

\begin{figure}
\begin{center}

\epsfig{figure=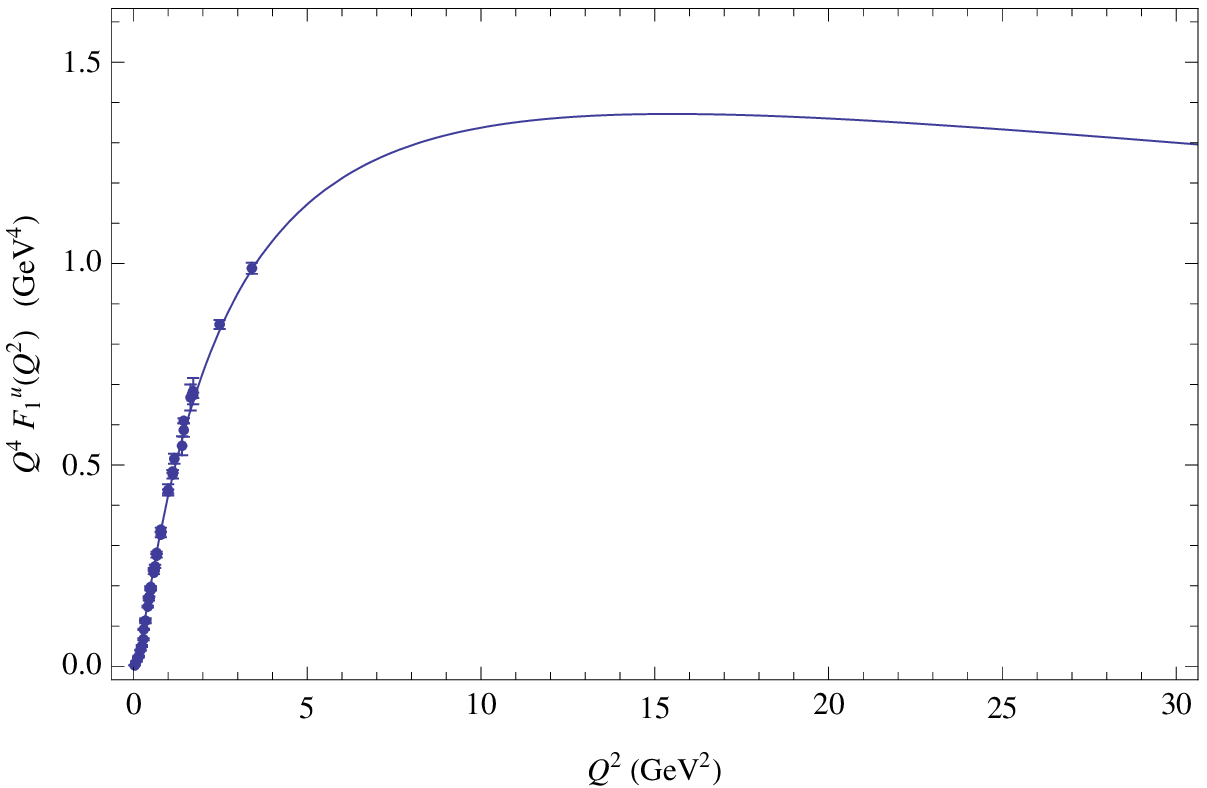,scale=.575}
\end{center}
\vspace*{-.6cm}
\noindent 
\caption{Dirac $u$ quark 
form factor multiplied by $Q^4$. 
\label{fig:F1uQ4}}  

\vspace*{-.35cm}

\begin{center}
\epsfig{figure=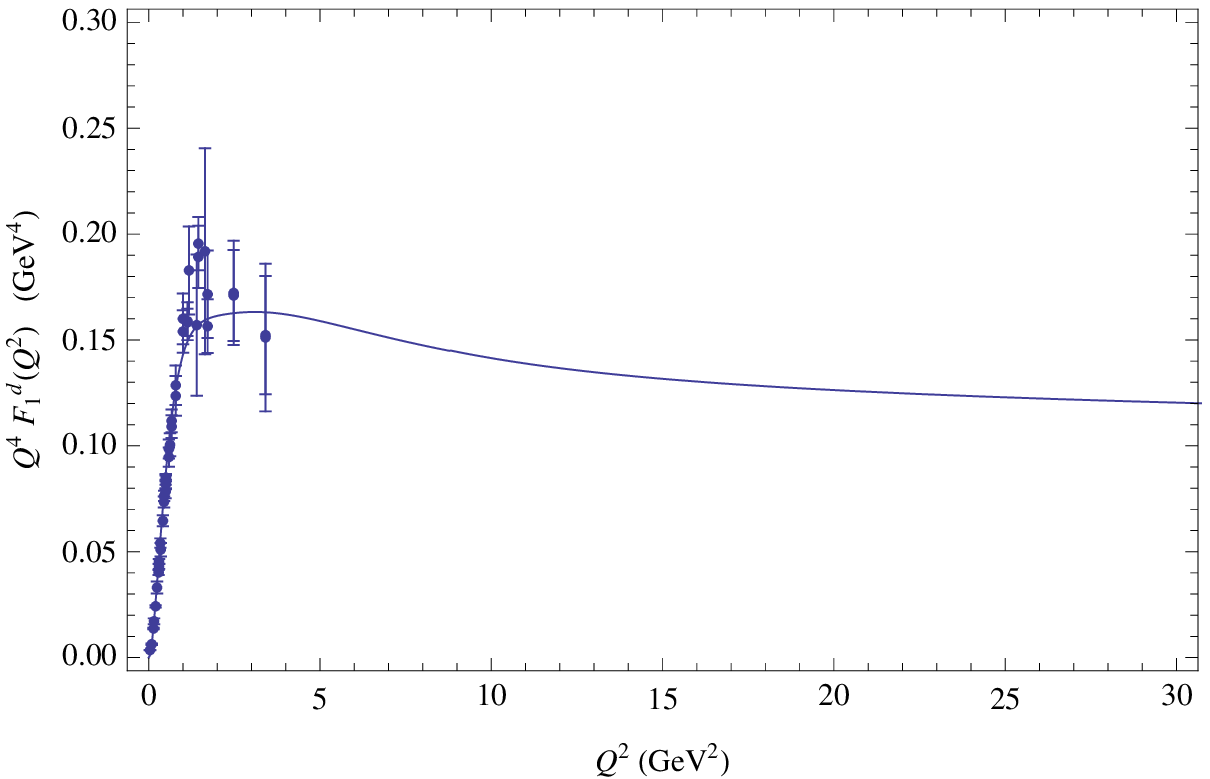,scale=.575}
\end{center}
\vspace*{-.6cm}
\noindent 
\caption{Dirac $d$ quark 
form factor multiplied by $Q^4$. 
\label{fig:F1dQ4}}

\vspace*{-.35cm}

\begin{center}
\epsfig{figure=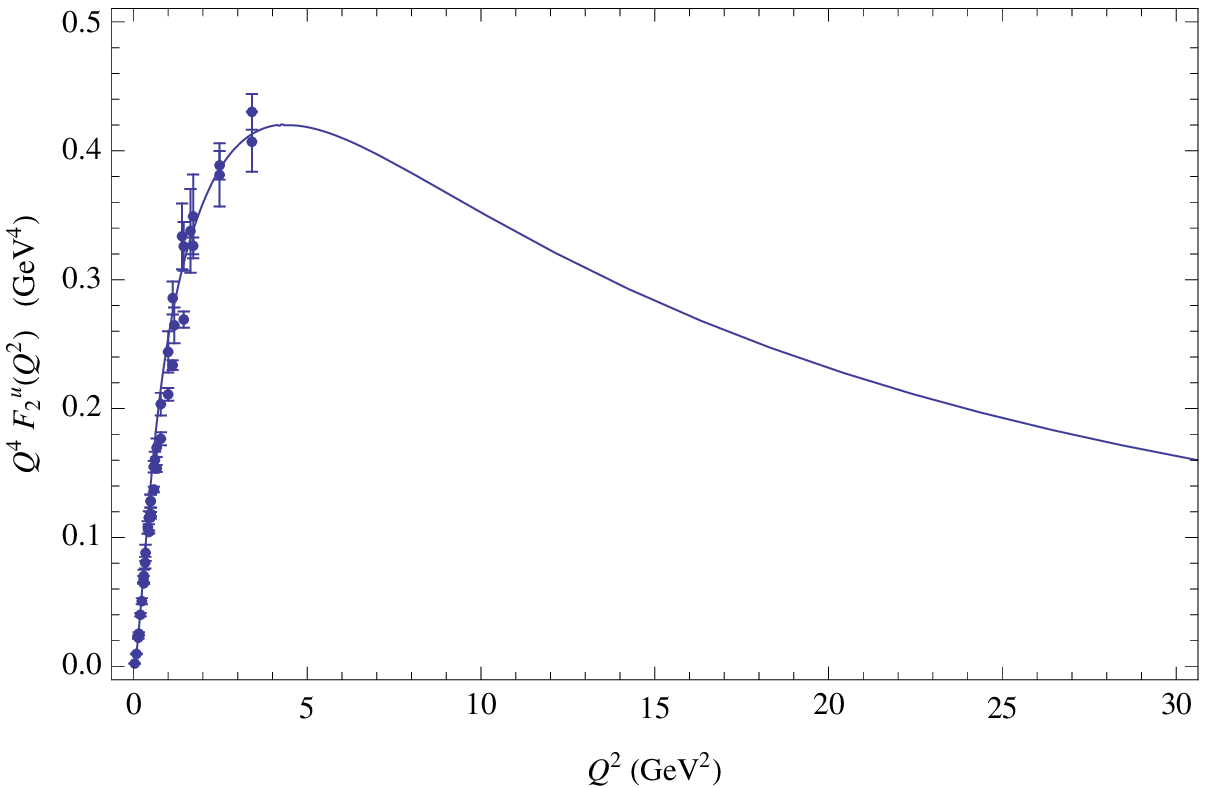,scale=.575}
\end{center}
\vspace*{-.6cm}
\noindent 
\caption{Pauli $u$ quark 
form factor multiplied by $Q^4$. 
\label{fig:F2uQ4}}  

\vspace*{-.35cm}

\begin{center}
\epsfig{figure=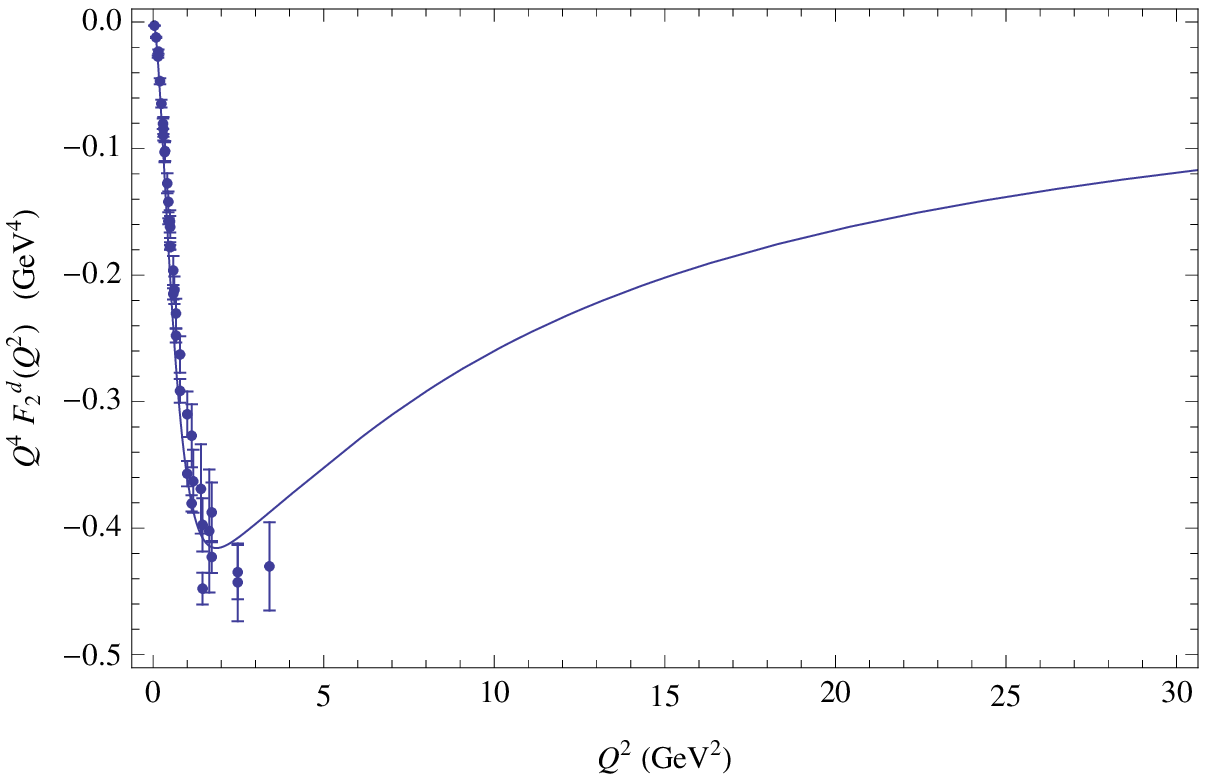,scale=.575}
\end{center}
\vspace*{-.6cm}
\noindent 
\caption{Pauli $d$ quark 
form factor multiplied by $Q^4$. 
\label{fig:F2dQ4}}  
\end{figure}

\begin{figure}
\begin{center}
\epsfig{figure=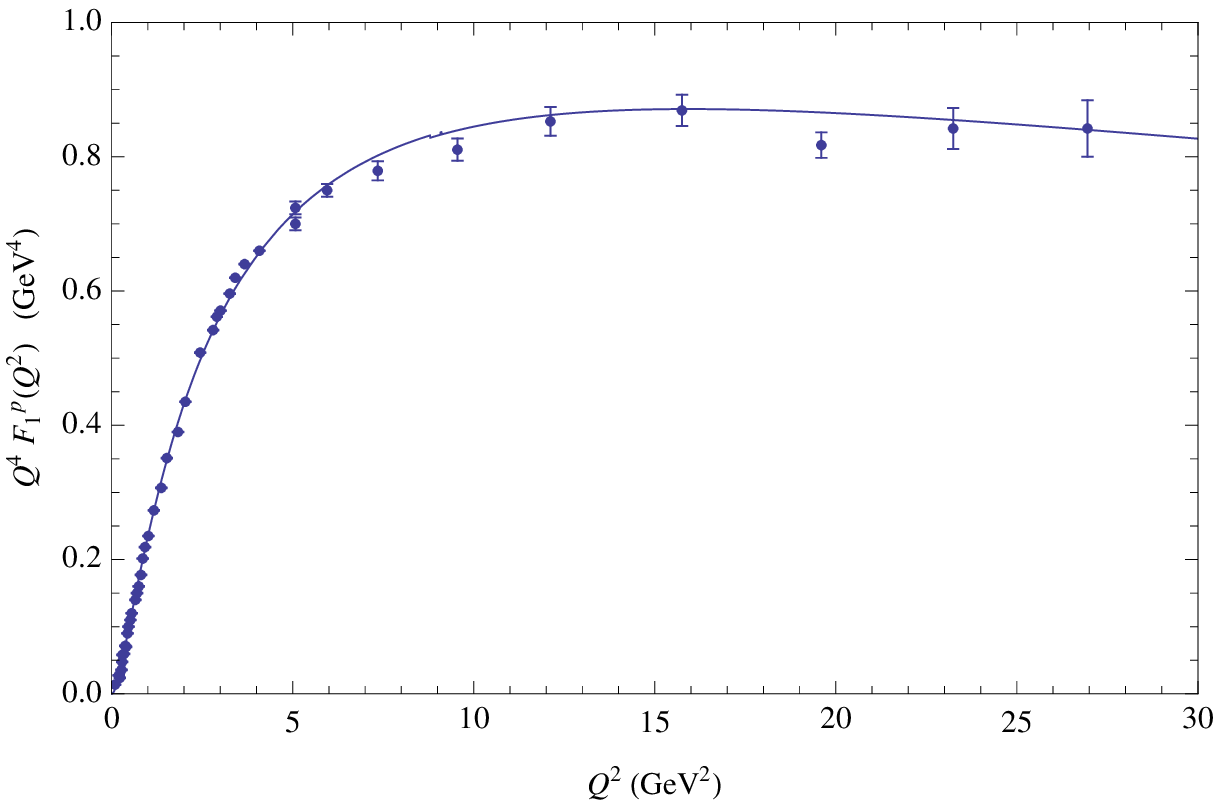,scale=.575}
\end{center}
\vspace*{-.6cm}
\noindent 
\caption{Dirac proton 
form factor multiplied by $Q^4$. 
\label{fig:F1pQ4}}  

\vspace*{-.35cm}

\begin{center}
\epsfig{figure=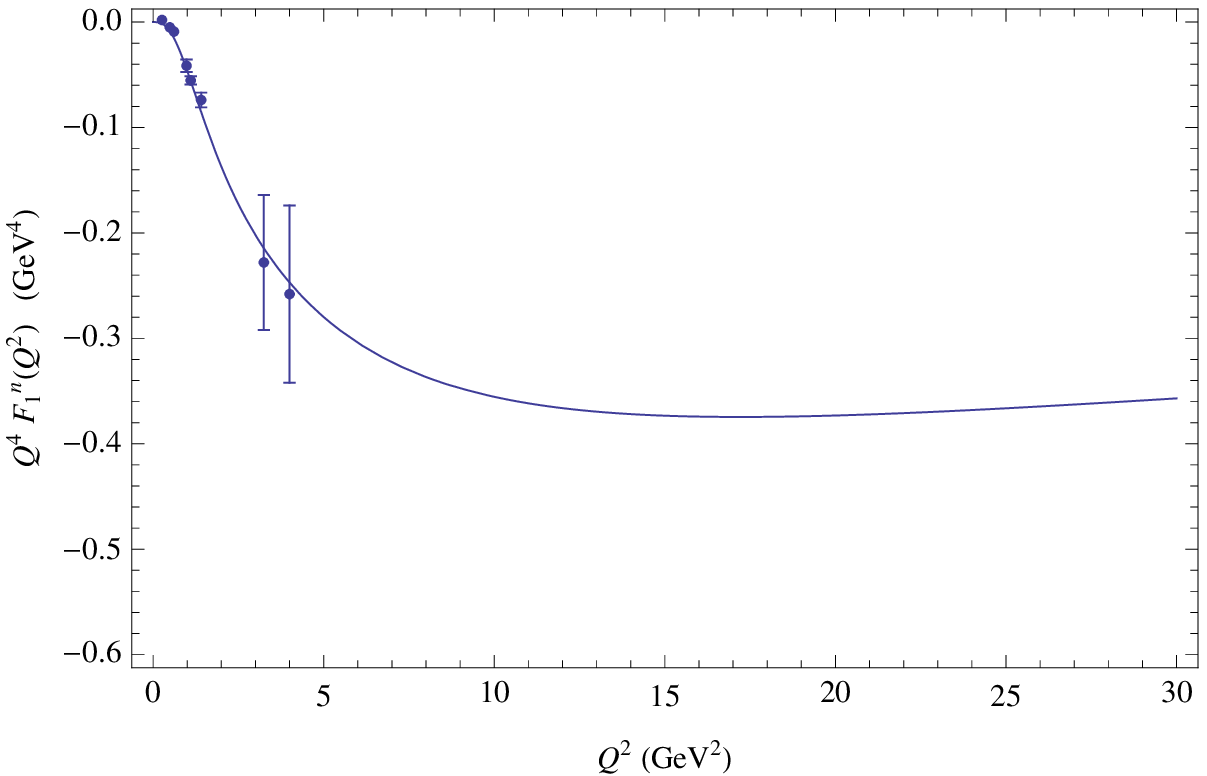,scale=.575}
\end{center}
\vspace*{-.6cm}
\noindent 
\caption{Dirac neutron 
form factor multiplied by $Q^4$. 
\label{fig:F1nQ4}}  

\vspace*{-.35cm}

\begin{center}
\epsfig{figure=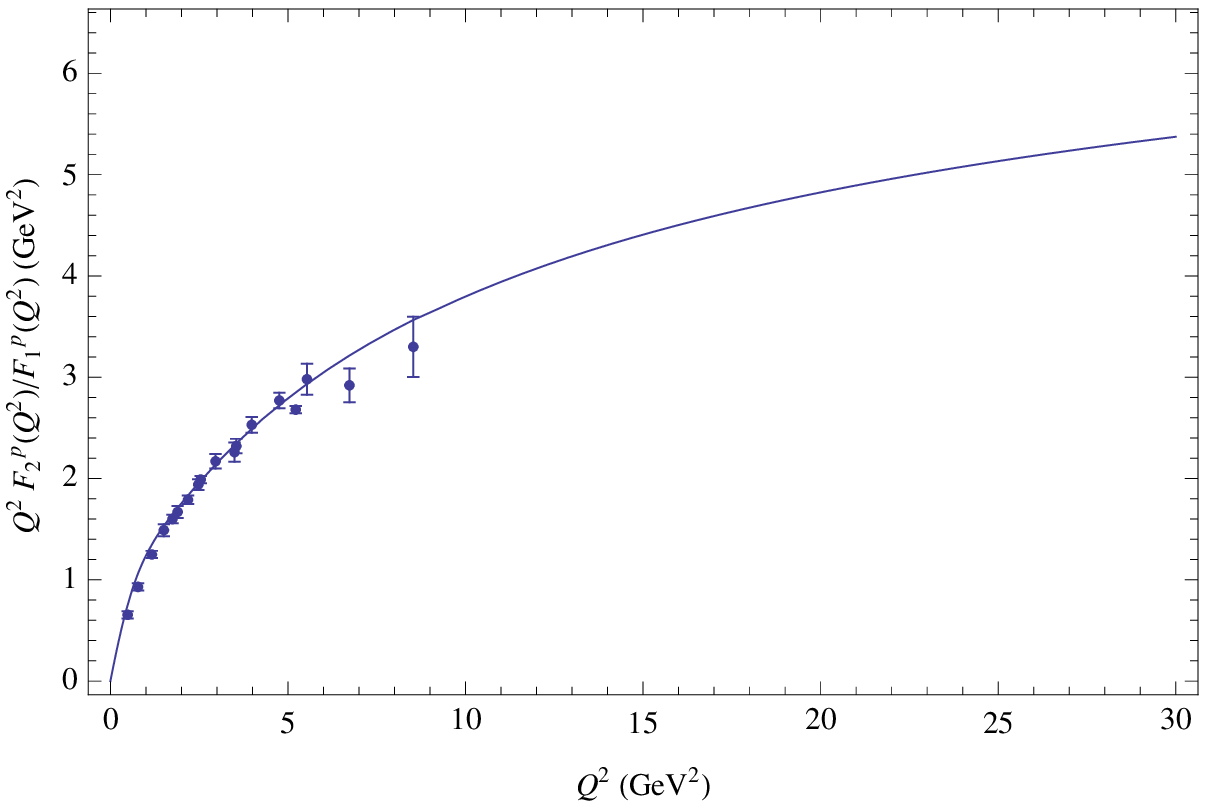,scale=.575}
\end{center}
\vspace*{-.6cm}
\noindent 
\caption{Ratio $Q^2 F_2^p(Q^2)/F_1^p(Q^2)$. 
\label{fig:F2pF1p}}  
\end{figure}

\begin{figure}
\begin{center}
\epsfig{figure=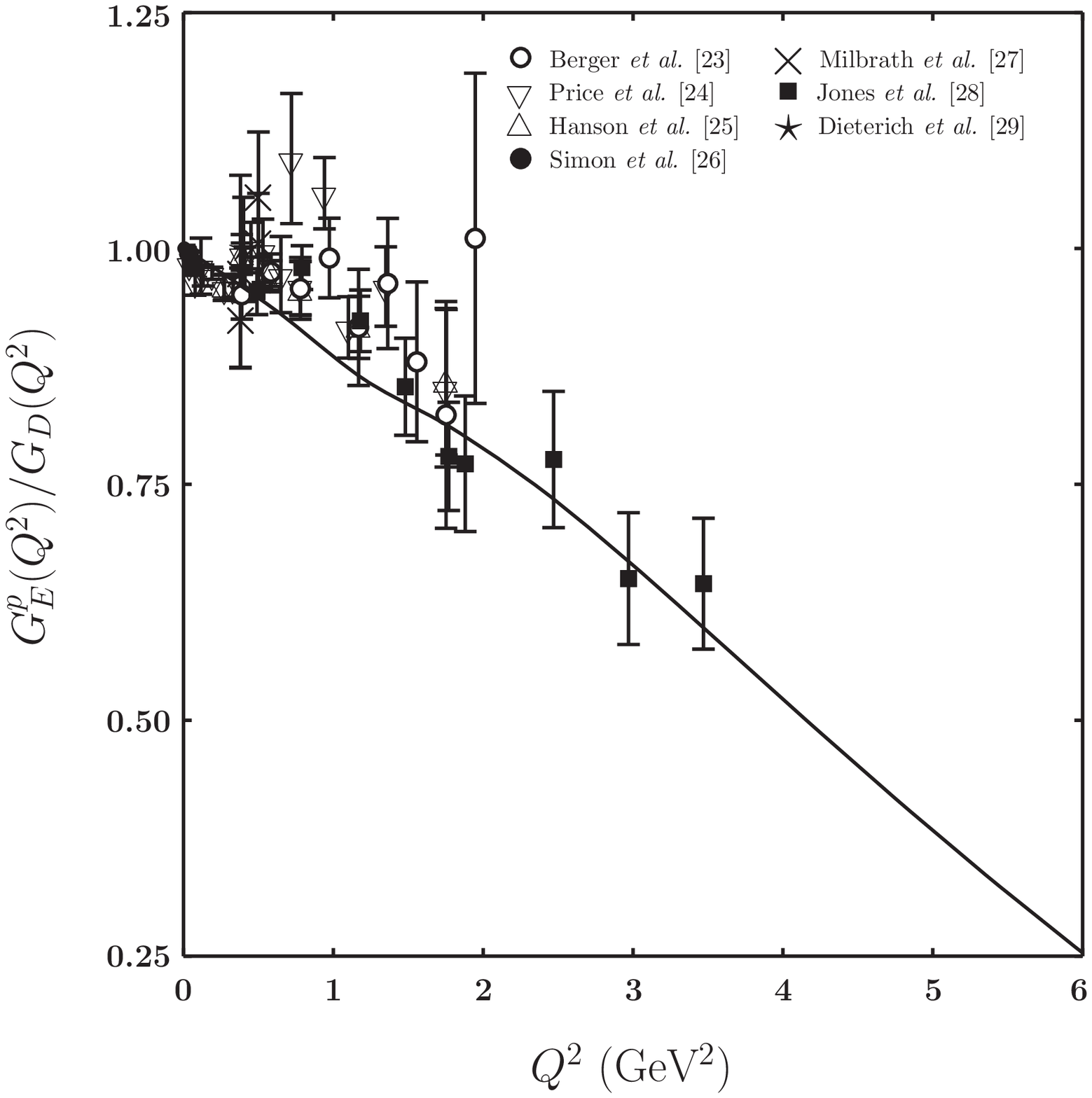,scale=.38}
\end{center}
\vspace*{-1.8cm}
\noindent 
\caption{Ratio $G_E^p(Q^2)/G_D(Q^2)$. 
\label{fig:gepd}}  

\begin{center}
\epsfig{figure=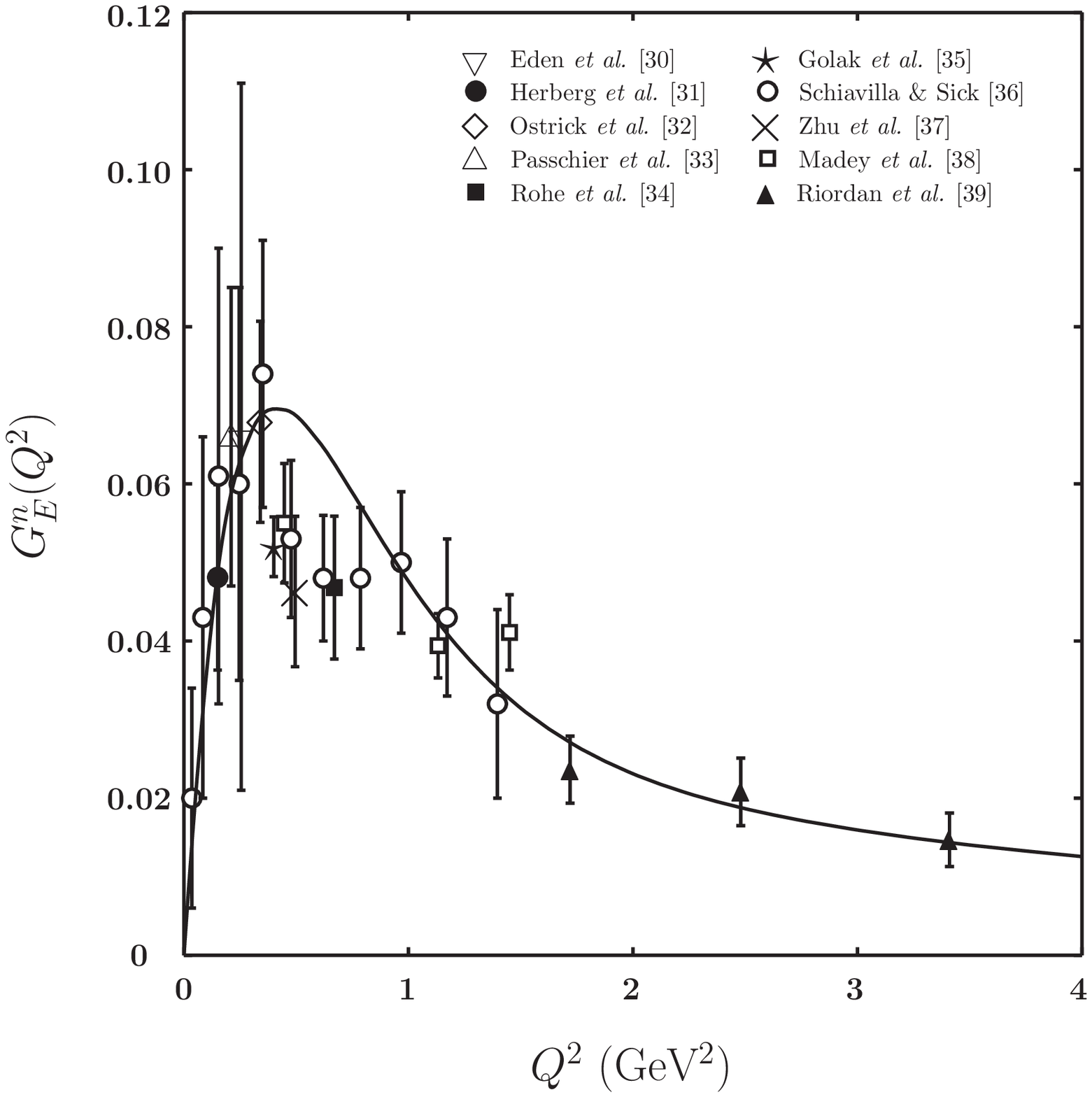,scale=.38}
\end{center}
\vspace*{-1.8cm}
\noindent 
\caption{Charge neutron form factor $G_E^n(Q^2)$.
\label{fig:gen}}  

\begin{center}
\epsfig{figure=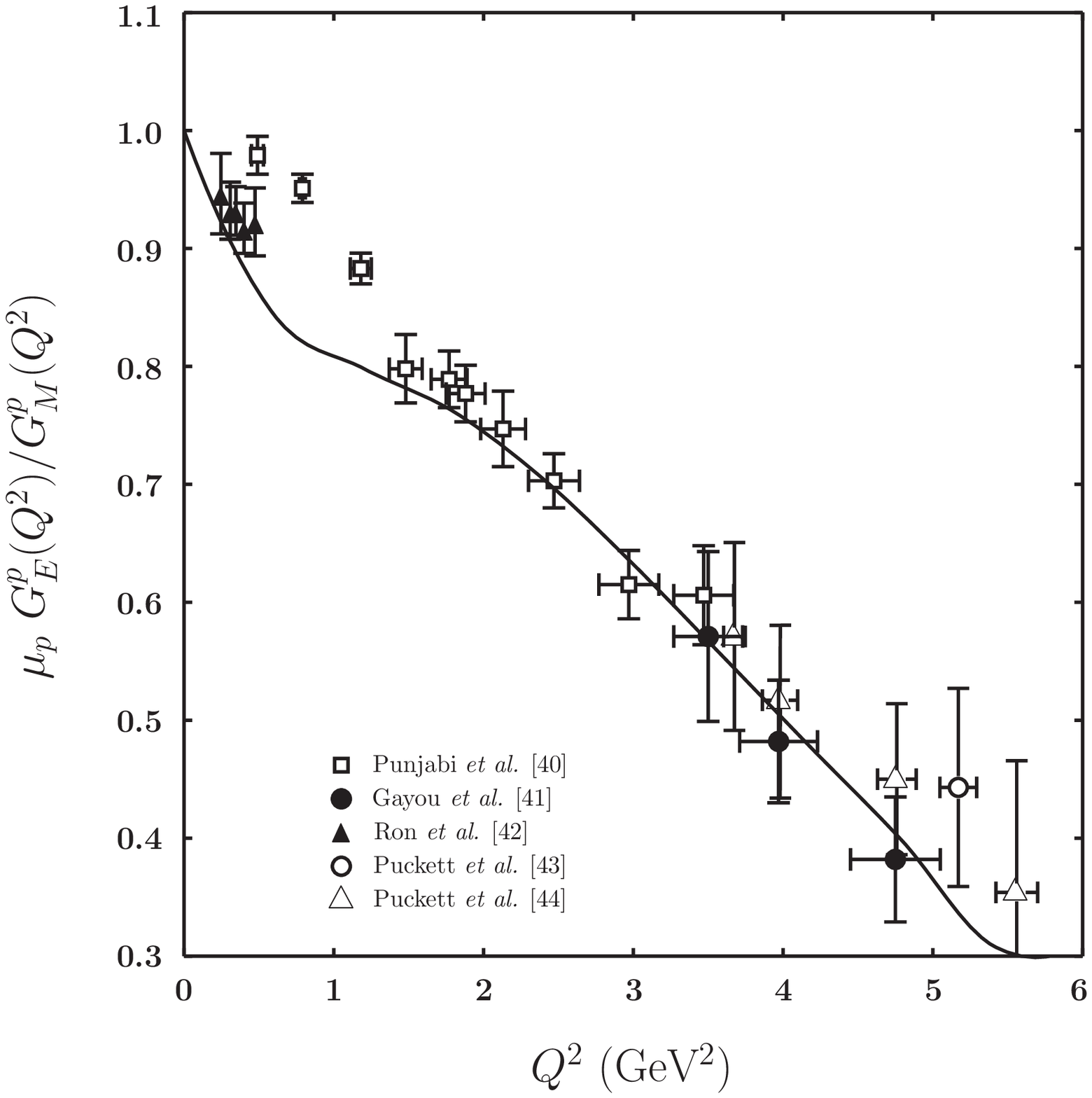,scale=.38}
\end{center}
\vspace*{-1.8cm}
\noindent 
\caption{Ratio $\mu_p G_E^p(Q^2)/G_M^p(Q^2)$. 
\label{fig:gepgmp}}  
\end{figure}

\begin{figure}
\begin{center}

\epsfig{figure=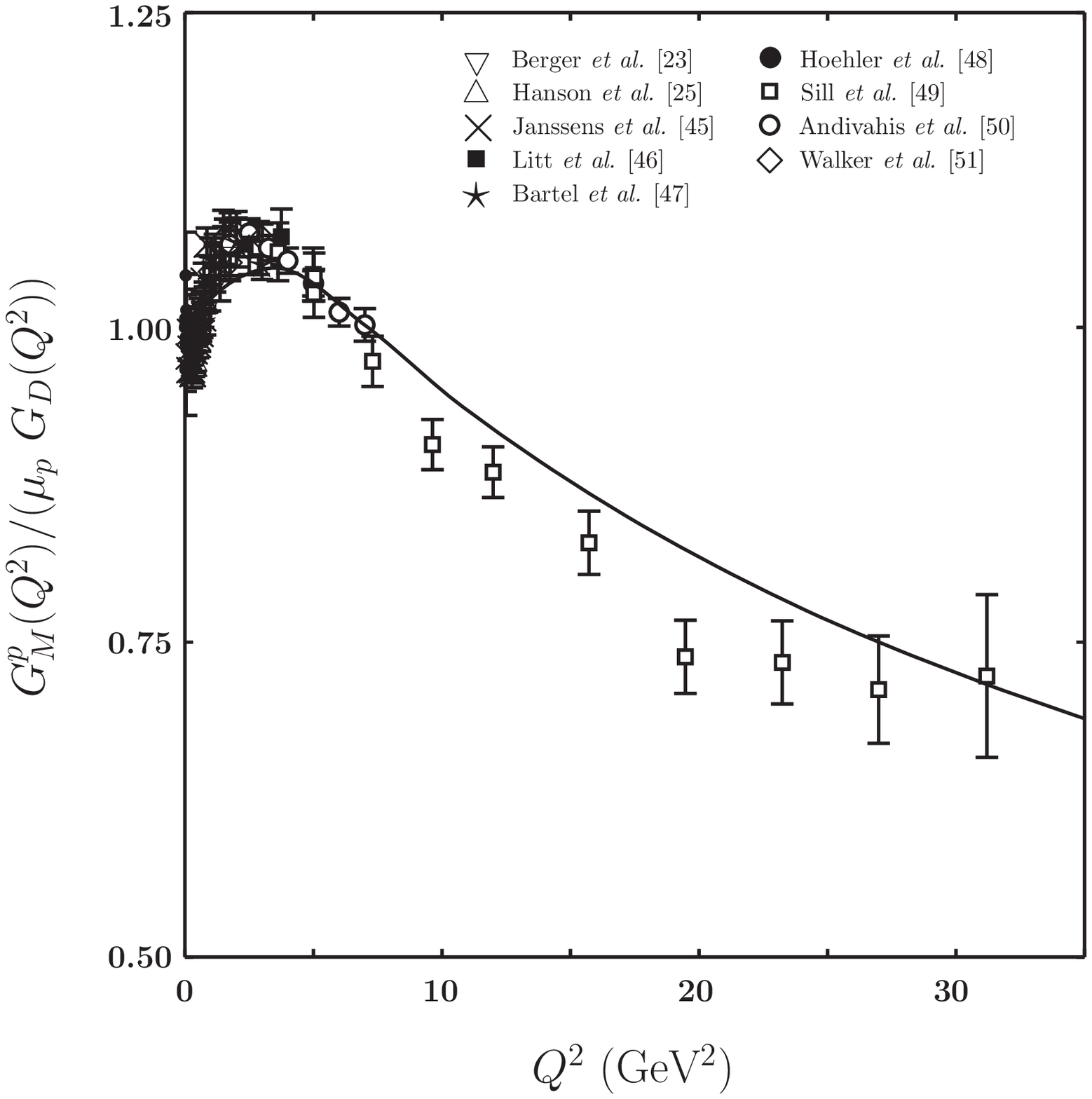,scale=.38}
\end{center}
\vspace*{-1.8cm}
\noindent 
\caption{Ratio $G_M^p(Q^2)/(\mu_p G_D(Q^2))$. 
\label{fig:gmpd}}

\begin{center}
\epsfig{figure=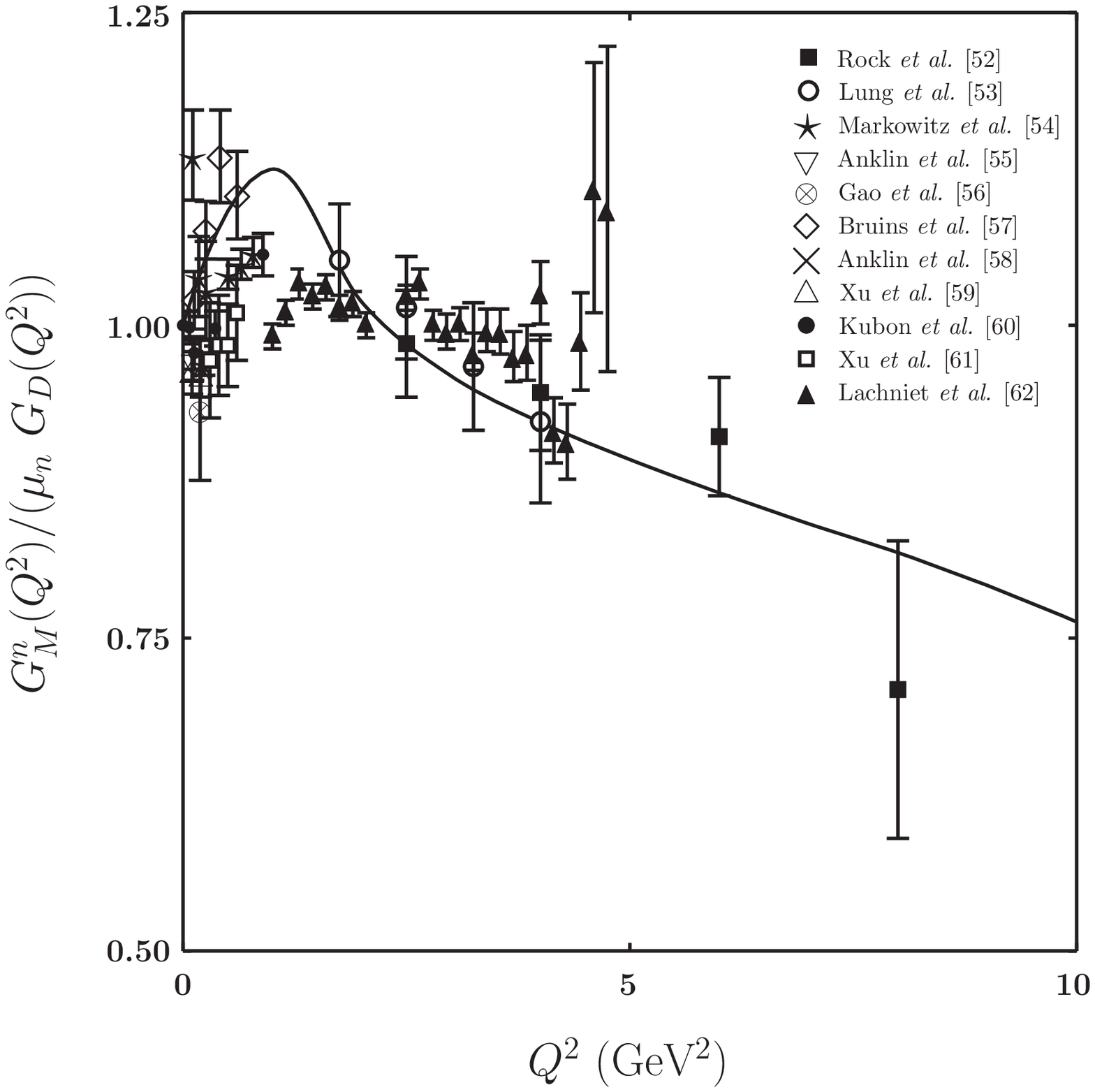,scale=.38}
\end{center}
\vspace*{-1.8cm}
\noindent 
\caption{Ratio $G_M^n(Q^2)/(\mu_n G_D(Q^2))$. 
\label{fig:gmnd}}  

\begin{center}
\epsfig{figure=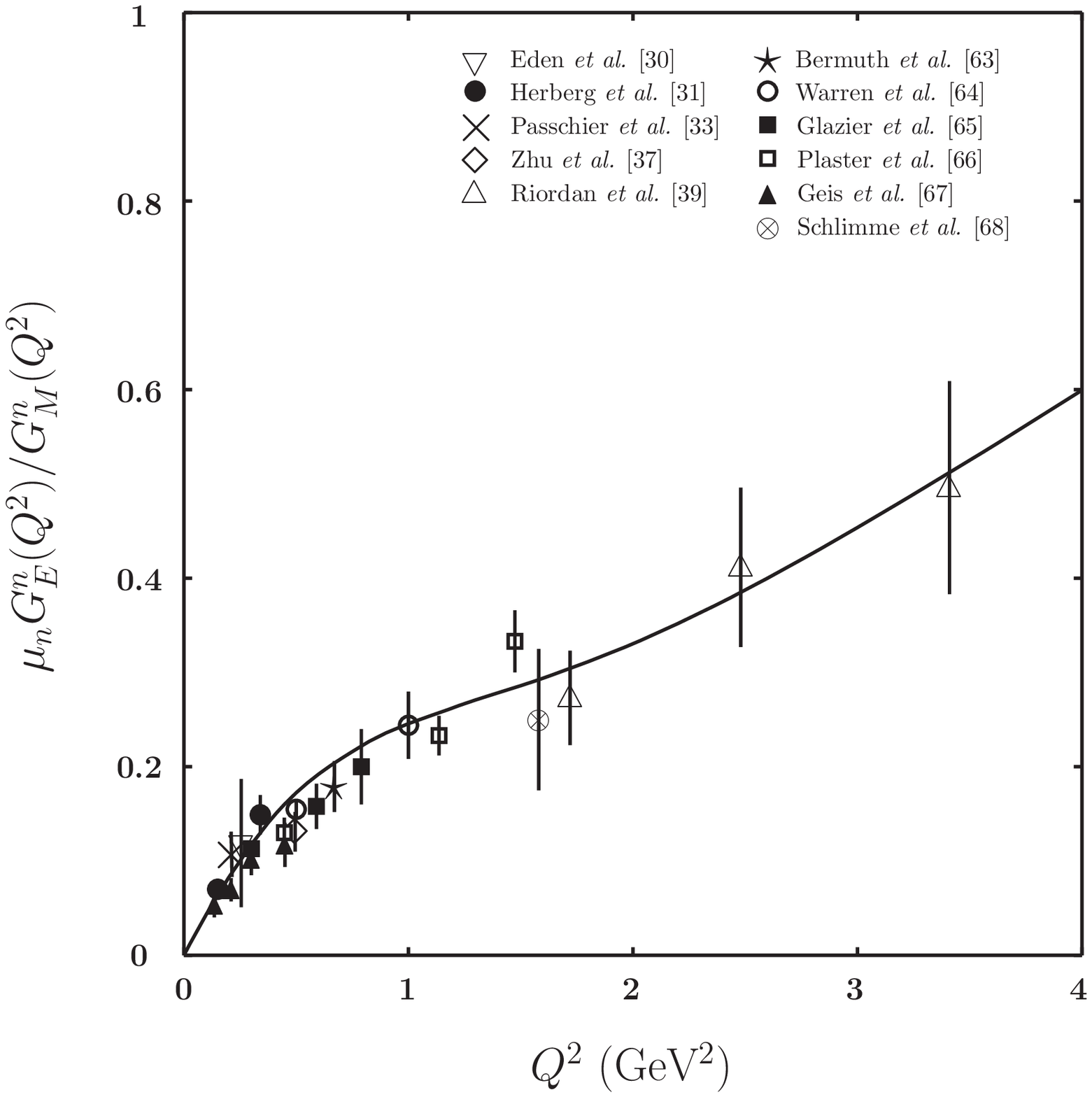,scale=.38}
\end{center}
\vspace*{-1.8cm}
\noindent 
\caption{Ratio $\mu_n G_E^n(Q^2)/G_M^n(Q^2)$. 
\label{fig:gengmn}}  
\end{figure}

Finally, we discuss our numerical results for 
electromagnetic properties of nucleons.  The fit results in values 
for the magnetic moments in terms of the nuclear magneton (n.m.) 
and for the electromagnetic radii as shown in Table~II, we also 
show the current data~\cite{Agashe:2014kda} on these quantities.   
In Figs.~\ref{fig:F1uQ4}-\ref{fig:F2pF1p} we present the plots of 
the Dirac and Pauli form factors for $u$ and $d$ quarks and nucleons. 
The data in Figs.~\ref{fig:F1uQ4}-\ref{fig:F2pF1p} 
are taken from Refs.~\cite{Cates:2011pz,Diehl:2013xca}.  
In Figs.~\ref{fig:gepd}-\ref{fig:gengmn} we also give 
results for the Sachs nucleon form factors and compare them with 
the dipole formula $G_D(Q^2) = 1/(1+Q^2/0.71 \ \mathrm{GeV}^2)^2$ 
and with data~\cite{berger71}-\cite{Schlimme:2013eoz}.  
Overall we have good agreement with the data. 

\begin{table}[htb]
\begin{center}
\caption{Electromagnetic properties of nucleons} 

\vspace*{.1cm}

\def\arraystretch{1.2}
    \begin{tabular}{|c|c|c|}
      \hline
Quantity & Our results & Data~\cite{Agashe:2014kda} \\
\hline
$\mu_p$ (in n.m.)          &  2.793       &  2.793              \\
\hline
$\mu_n$ (in n.m.)          & -1.913       & -1.913              \\
\hline
$r_E^p$ (fm)     &  0.781 &  0.84087 $\pm$ 0.00026 $\pm$ 0.00029 \\ 
\hline
$\la r^2_E \ra^n$ (fm$^2$) & -0.112 & -0.1161 $\pm$ 0.0022 \\ 
\hline
$r_M^p$ (fm)     &  0.717 &  0.777  $\pm$ 0.013 $\pm$ 0.010 \\ 
\hline
$r_M^n$ (fm)     &  0.694 &  0.862$^{+0.009}_{-0.008}$     \\
\hline
\end{tabular}
\end{center}
\end{table}

\section{Conclusion} 

In conclusion, we want to summarize the main result of our paper. 
Using global fits of valence $u$ and $d$ quark parton distributions 
and data on quark and nucleon form factors in the Euclidean region, 
we construct a light-front quark model for the nucleon structure  
consistent with model-independent scaling laws --- 
the Drell-Yan-West duality~\cite{Drell:1969km} and 
quark counting rules~\cite{Brodsky:1973kr}.

\begin{acknowledgments}

The authors thank Werner Vogelsang for the useful discussions. 
This work was supported by the DFG under Contract No. LY 114/2-1, 
by the German Bundesministerium f\"ur Bildung und Forschung (BMBF) 
under Grant No. 05P12VTCTG, 
by Marie Curie Reintegration Grant No. IRG 256574, 
by CONICYT (Chile) Research Project No. 80140097,  
by CONICYT (Chile) under Grant No. 7912010025, 
by FONDECYT (Chile) under Grants No. 1140390 and No. 1141280,    
and by the Tomsk State University 
Competitiveness Improvement Program. 
V.E.L. would like to thank the Departamento de F\'\i sica y Centro
Cient\'\i fico Tecnol\'ogico de Valpara\'\i so (CCTVal), Universidad
T\'ecnica Federico Santa Mar\'\i a, Valpara\'\i so, Chile for the warm
hospitality. 

\end{acknowledgments} 

\clearpage 

\clearpage 

\clearpage 

\clearpage 

\begin{thebibliography}{99}

\bibitem{Gutsche:2013zia} 
  T.~Gutsche, V.~E.~Lyubovitskij, I.~Schmidt and A.~Vega,
  Phys.\ Rev.\ D {\bf 89}, 054033 (2014). 

\bibitem{Brodsky:1973kr} 
  S.~J.~Brodsky and G.~R.~Farrar,
  Phys.\ Rev.\ Lett.\  {\bf 31}, 1153 (1973); 
  V.~A.~Matveev, R.~M.~Muradian and A.~N.~Tavkhelidze,
  Lett.\ Nuovo Cim.\  {\bf 7}, 719 (1973).  

\bibitem{Martin:2009iq} 
  A.~D.~Martin, W.~J.~Stirling, R.~S.~Thorne and G.~Watt,
  Eur.\ Phys.\ J.\ C {\bf 63}, 189 (2009). 

\bibitem{Aicher:2010cb} 
  M.~Aicher, A.~Schafer and W.~Vogelsang,
  Phys.\ Rev.\ Lett.\  {\bf 105}, 252003 (2010). 

\bibitem{Gutsche:2014zua} 
  T.~Gutsche, V.~E.~Lyubovitskij, I.~Schmidt and A.~Vega,
  arXiv:1410.6424 [hep-ph].


\bibitem{Cates:2011pz}
  G.~D.~Cates, C.~W.~de Jager, S.~Riordan and B.~Wojtsekhowski,
  Phys.\ Rev.\ Lett.\  {\bf 106}, 252003 (2011). 

\bibitem{Diehl:2013xca} 
  M.~Diehl and P.~Kroll,
  Eur.\ Phys.\ J.\ C {\bf 73}, 2397 (2013); 
  M.~Diehl,
  Nucl.\ Phys.\ Proc.\ Suppl.\  {\bf 161}, 49 (2006). 

\bibitem{Obukhovsky:2013fpa} 
  I.~T.~Obukhovsky, A.~Faessler, T.~Gutsche and \\{} 
  V.~E.~Lyubovitskij,
  Phys.\ Rev.\ D {\bf 89}, 014032 (2014); \\{}
  J.\ Phys.\ G {\bf 41}, 095005 (2014).

\bibitem{Diehl:2004cx}
  M.~Diehl, T.~Feldmann, R.~Jakob and P.~Kroll,
  Eur.\ Phys.\ J.\ C {\bf 39}, 1 (2005);
  M.~Guidal, M.~V.~Polyakov, A.~V.~Radyushkin and M.~Van\-der\-hae\-ghen,
  Phys.\ Rev.\ D {\bf 72}, 054013 (2005);
  O.~V.~Selyugin and O.~V.~Teryaev,
  Phys.\ Rev.\ D {\bf 79}, 033003 (2009).

\bibitem{Radyushkin:1998rt} 
  A.~V.~Radyushkin,
  Phys.\ Rev.\ D {\bf 58}, 114008 (1998). 

\bibitem{Mueller:1998fv}
  D.~Mueller, D.~Robaschik, B.~Geyer, F.~M.~Dittes and J.~Horejsi,
  Fortsch.\ Phys.\  {\bf 42}, 101 (1994); 
  X.~D.~Ji,
  Phys.\ Rev.\ Lett.\  {\bf 78}, 610 (1997);
  A.~V.~Radyushkin,
  Phys.\ Rev.\ D {\bf 56}, 5524 (1997). 

\bibitem{Brodsky_Drell} 
  S.~J.~Brodsky, S.~D.~Drell,
  Phys.\ Rev.\  D {\bf 22}, 2236 (1980).  

\bibitem{LFQCD}
  S.~J.~Brodsky and D.~S.~Hwang,
  Nucl.\ Phys.\ B {\bf 543}, 239 (1999); 
  S.~J.~Brodsky, D.~S.~Hwang, B.~-Q.~Ma and I.~Schmidt,
  Nucl.\ Phys.\ B {\bf 593}, 311 (2001); 
  S.~J.~Brodsky, M.~Diehl and D.~S.~Hwang,
  Nucl.\ Phys.\ B {\bf 596}, 99 (2001). 

\bibitem{Brodsky:2007hb}
  S.~J.~Brodsky, G.~F.~de Teramond,
  Phys.\ Rev.\  D {\bf 77}, 056007 (2008); 
  G.~F.~de Teramond and S.~J.~Brodsky,
  AIP Conf.\ Proc.\  {\bf 1432}, 168 (2012); 
  S.~J.~Brodsky and G.~F.~de Teramond,
  AIP Conf.\ Proc.\  {\bf 1388}, 22 (2011). 

\bibitem{Brodsky:2011xx} 
  S.~J.~Brodsky, F.~-G.~Cao and G.~F.~de Teramond,
  Phys.\ Rev.\ D {\bf 84}, 075012 (2011). 

\bibitem{Abidin:2009hr}
  Z.~Abidin and C.~E.~Carlson,
  Phys.\ Rev.\ D {\bf 79}, 115003 (2009). 

\bibitem{Vega:2009zb} 
  A.~Vega, I.~Schmidt, T.~Branz, T.~Gutsche and \\{} V.~E.~Lyubovitskij,
  Phys.\ Rev.\ D {\bf 80}, 055014 (2009); 
  T.~Branz, T.~Gutsche, 
  V.~E.~Lyubovitskij, I.~Schmidt, A.~Vega,
  Phys.\ Rev.\  D {\bf 82}, 074022 (2010); 
  T.~Gutsche, V.~E.~Lyubovitskij, I.~Schmidt and A.~Vega,
  Phys.\  Rev.\ D {\bf 87}, 056001 (2013). 

\bibitem{Gutsche:2011vb}
  T.~Gutsche, V.~E.~Lyubovitskij, I.~Schmidt and A.~Vega, \\{}
  Phys.\ Rev.\ D {\bf 85}, 076003 (2012); 
  Phys.\ Rev.\ D {\bf 86}, 036007 (2012); 
  Phys.\ Rev.\ D {\bf 87}, 016017 (2013). 

\bibitem{Brodsky:1994kg}
  S.~J.~Brodsky, M.~Burkardt and I.~Schmidt,
  Nucl.\ Phys.\ B {\bf 441}, 197 (1995). 

\bibitem{Yuan:2003fs}
  F.~Yuan,
  Phys.\ Rev.\ D {\bf 69}, 051501 (2004). 

\bibitem{Drell:1969km} 
  S.~D.~Drell and T.~-M.~Yan,
  Phys.\ Rev.\ Lett.\  {\bf 24}, 181 (1970); 
  G.~B.~West,
  Phys.\ Rev.\ Lett.\  {\bf 24}, 1206 (1970).

\bibitem{Agashe:2014kda}
  K.~A.~Olive {\it et al.}  (Particle Data Group Collaboration),
  Chin.\ Phys.\ C {\bf 38}, 090001 (2014).

\bibitem{berger71}
  C.~Berger, V.~Burkert, G.~Knop, B.~Langenbeck and K.~Rith,
  Phys.\ Lett.\ B {\bf 35}, 87 (1971).

\bibitem{price71}
  L.~E.~Price {\it et al.}, 
  Phys.\ Rev.\ D {\bf 4}, 45 (1971).

\bibitem{hanson73}
  K.~M.~Hanson {\it et al.}, 
  Phys.\ Rev.\ D {\bf 8}, 753 (1973).

\bibitem{simon80}
  G.~G.~Simon, C.~Schmitt, F.~Borkowski and \\{} V.~H.~Walther,
  Nucl.\ Phys.\ A {\bf 333}, 381 (1980).

\bibitem{milbrath98}
 B.~D.~Milbrath {\it et al.} (Bates FPP Collaboration), 
 Phys.\ Rev.\ Lett.\  {\bf 80}, 452 (1998); 
                      {\bf 82}, 2221(E) (1999).  


\bibitem{jones00}
  M.~K.~Jones {\it et al.} (Jefferson Lab Hall A Collaboration),
  Phys.\ Rev.\ Lett.\  {\bf 84}, 1398 (2000).    

\bibitem{dieterich01}
 S. Dieterich {\it et al.}, 
  Phys.\ Lett.\ B {\bf 500}, 47 (2001).  

\bibitem{eden94}
 T.~Eden {\it et al.}, 
Phys.\ Rev.\  C {\bf 50}, R1749 (1994).

\bibitem{herberg99}
 C.~Herberg {\it et al.},
  Eur.\ Phys.\ J.\ A {\bf 5}, 131 (1999).

\bibitem{ostrick99}
 M.~Ostrick {\it et al.}, 
  Phys.\ Rev.\ Lett.\  {\bf 83}, 276 (1999).

\bibitem{passchier99}
 I.~Passchier {\it et al.}, 
  Phys.\ Rev.\ Lett.\  {\bf 82}, 4988 (1999).  

\bibitem{rohe99}
 D. Rohe {\it et al.}, 
  Phys.\ Rev.\ Lett.\  {\bf 83}, 4257 (1999).

\bibitem{golak01}
J.~Golak, G.~Ziemer, H.~Kamada, H.~Witala and \\{}W.~Gl\"ockle,
  Phys.\ Rev.\ C {\bf 63}, 034006 (2001). 

\bibitem{schiavilla01}
R.~Schiavilla and I.~Sick,
  Phys.\ Rev.\ C {\bf 64}, 041002 (2001). 

\bibitem{zhu01}
 H.~Zhu {\it et al.} (Jefferson Lab E93-026 Collaboration),
  Phys.\ Rev.\ Lett.\  {\bf 87}, 081801 (2001). 

\bibitem{madey03}
 R. Madey {\it et al.} (Jefferson Lab E93-038 Collaboration),
  Phys.\ Rev.\ Lett.\  {\bf 91}, 122002 (2003). 

\bibitem{Riordan:2010id} 
  S.~Riordan {\it et al.},
  Phys.\ Rev.\ Lett.\  {\bf 105}, 262302 (2010). 


\bibitem{punjabi05}
   V.~Punjabi {\it et al.}, 
  Phys.\ Rev.\ C {\bf 71}, 055202 (2005); 
  C {\bf 71}, 069902(E) (2005).  

\bibitem{gayou02}
 O.~Gayou {\it et al.}  (Jefferson Lab Hall A Collaboration), 
  Phys.\ Rev.\ Lett.\  {\bf 88}, 092301 (2002). 
 
\bibitem{ron11}
 G.~Ron {\it et al.} (JLab Hall A Collaboration),
  Phys.\ Rev.\ C {\bf 84}, 055204 (2011). 

\bibitem{puckett10}
 A.~J.~R.~Puckett {\it et al.}, 
  Phys.\ Rev.\ Lett.\  {\bf 104}, 242301 (2010). 

\bibitem{puckett12}
 A.~J.~R.~Puckett {\it et al.}, 
  Phys.\ Rev.\ C {\bf 85}, 045203 (2012). 

\bibitem{Janssens:1966}
  T.~Janssens, R.~Hofstadter, E.~B.~Hughes and M.~R.~Yearian,
  Phys.\ Rev.\ {\bf 142}, 922 (1966).
\bibitem{Litt:1969my}
  J.~Litt {\it et al.},
  Phys.\ Lett.\ B {\bf 31}, 40 (1970).

\bibitem{bartel73}
 W. Bartel {\it et al.}, 
  Nucl.\ Phys.\  {\bf B58}, 429 (1973).

\bibitem{Hohler:1976ax}
  G.~Hohler {\it et al.}, 
  Nucl.\ Phys.\ B {\bf 114}, 505 (1976).

\bibitem{Sill:1992qw}
  A.~F.~Sill {\it et al.},
  Phys.\ Rev.\ D {\bf 48}, 29 (1993).

\bibitem{Andivahis:1994rq}
  L.~Andivahis {\it et al.},
  Phys.\ Rev.\ D {\bf 50}, 5491 (1994).

\bibitem{Walker:1993vj}
  R.~C.~Walker {\it et al.},
  Phys.\ Rev.\ D {\bf 49} 5671 (1994).

\bibitem{Rock:1982gf}
  S.~Rock {\it et al.},
  Phys.\ Rev.\ Lett.\  {\bf 49}, 1139 (1982).

\bibitem{Lung:1992bu}
  A.~Lung {\it et al.},
  Phys.\ Rev.\ Lett.\  {\bf 70}, 718 (1993).

\bibitem{Markowitz:1993hx}
  P.~Markowitz {\it et al.},
  Phys.\ Rev.\ C {\bf 48}, R5 (1993).

\bibitem{Anklin:1994ae}
  H.~Anklin {\it et al.},
  Phys.\ Lett.\ B {\bf 336}, 313 (1994).


\bibitem{Gao:1994ud}
  H.~Gao {\it et al.},
  Phys.\ Rev.\ C {\bf 50}, R546 (1994). 

\bibitem{Bruins:1995ns}
  E.~E.~W.~Bruins {\it et al.},
  Phys.\ Rev.\ Lett.\  {\bf 75}, 21 (1995).

\bibitem{Anklin:1998ae}
  H.~Anklin {\it et al.},
  Phys.\ Lett.\ B {\bf 428}, 248 (1998).

\bibitem{xu00}
  W.~Xu {\it et al.} 
  Phys.\ Rev.\ Lett.\  {\bf 85}, 2900 (2000). 

\bibitem{Kubon:2001rj}
  G.~Kubon {\it et al.},
  Phys.\ Lett.\ B {\bf 524}, 26 (2002). 

\bibitem{Xu:2002xc}
  W.~Xu {\it et al.} (Jefferson Lab E95-001 Collaboration),
  Phys.\ Rev.\ C {\bf 67}, 012201 (2003). 

\bibitem{Lachniet:2008qf}
  V.~J.~Lachniet {\it et al.} (CLAS Collaboration), 
  Phys.\ Rev.\ Lett.\  {\bf 102}, 192001 (2009). 

\bibitem{Bermuth:2003qh}
  J.~Bermuth {\it et al.},
  Phys.\ Lett.\ B {\bf 564}, 199 (2003). 

\bibitem{Warren:2003ma}
  G.~Warren {\it et al.}  (Jefferson Laboratory E93-026 Collaboration),
  Phys.\ Rev.\ Lett.\  {\bf 92}, 042301 (2004). 

\bibitem{Glazier:2004ny}
  D.~I.~Glazier {\it et al.},
  Eur.\ Phys.\ J.\ A {\bf 24}, 101 (2005)

\bibitem{Plaster:2005cx}
  B.~Plaster {\it et al.}  (Jefferson Laboratory E93-038 Collaboration),
  Phys.\ Rev.\  C {\bf 73}, 025205 (2006). 

\bibitem{Geis:2008aa}
  E.~Geis {\it et al.}  (BLAST Collaboration),
  Phys.\ Rev.\ Lett.\  {\bf 101}, 042501 (2008). 

\bibitem{Schlimme:2013eoz} 
  B.~S.~Schlimme {\it et al.},
  Phys.\ Rev.\ Lett.\  {\bf 111}, 132504 (2013). 


\end{thebibliography}
\end{document}